\documentclass[apj,showkeys]{emulateapj}
\usepackage{amsmath, amsthm, amssymb}

\begin{document}

\title{How Good a Clock is Rotation?
The Stellar Rotation-Mass-Age Relationship for Old Field Stars}
\shorttitle{How Good a Clock is Rotation?}
\shortauthors{Epstein \& Pinsonneault}
\author{
Courtney~R.~Epstein\altaffilmark{1} and
Marc~H.~Pinsonneault\altaffilmark{1}
}
\altaffiltext{1}{Department of Astronomy, Ohio State University,
140 W.\ 18th Ave., Columbus, OH 43210, USA;
epstein,pinsono@astronomy.ohio-state.edu}

 \begin{abstract}
The rotation-mass-age relationship offers a promising avenue for measuring the ages of field stars, assuming the attendant uncertainties to this technique can be well characterized. We model stellar angular momentum evolution starting with a rotation distribution from open cluster M37. Our predicted rotation-mass-age relationship shows significant zero-point offsets compared to an alternative angular momentum loss law and published gyrochronology relations. Systematic errors at the 30 percent level are permitted by current data, highlighting the need for empirical guidance. We identify two fundamental sources of uncertainty that limit the precision of rotation-based ages and quantify their impact. Stars are born with a range of rotation rates, which leads to an age range at fixed rotation period. We find that the inherent ambiguity from the initial conditions is important for all young stars, and remains large for old stars below 0.6 M$_\odot$. Latitudinal surface differential rotation also introduces a minimum uncertainty into rotation period measurements and, by extension, rotation-based ages. Both models and the data from binary star systems 61 Cyg and $\alpha$ Cen demonstrate that latitudinal differential rotation is the limiting factor for rotation-based age precision among old field stars, inducing uncertainties at the $\sim2$ Gyr level. We also examine the relationship between variability amplitude, rotation period, and age. Existing ground-based surveys can detect field populations with ages as old as 1 -- 2 Gyr, while space missions can detect stars as old as the Galactic disk. In comparison with other techniques for measuring the ages of lower main sequence stars, including geometric parallax and asteroseismology, rotation-based ages have the potential to be the most precise chronometer for 0.6 -- 1.0 M$_\odot$ stars.
\end{abstract}

\section{Introduction}

Ages of stars are both fundamental and difficult to measure. Stars of known age, such as members of star clusters, have therefore been studied intensively. For the far more numerous field stars, by contrast, our tools for inferring ages are blunt and inexact (see \citealt{Soderblom 2010} for a nice review of techniques used to estimate ages). We can infer ages for some binary stars, and also for turnoff stars, from the combination of parallax and metallicity measurements. However, for typical GKM dwarfs, slow luminosity evolution on the main sequence and ambiguities from metallicity and helium errors result in uncertain age estimates. Measurements of Ca \textsc{II} H and K emission for main sequence stars led \citet{Wilson 1963} to insightfully notice an inverse relationship between chromospheric activity and stellar age. In an influential paper, \citet{Skumanich 1972} argued that stellar rotation, activity, and lithium abundances for solar-like stars all decrease as the square root of age. Rotation therefore provides an attractive alternative chronometer (e.g.\ \citealt{Lachaume 1999,Barnes 2007}). Stars with minimal \emph{nuclear} evolution can experience strong \emph{rotational} evolution, in the sense that low mass stars are observed to rotate more slowly as they age. The idea of rotation as a clock for low mass stars is an old one; however, the prospect of rotation as a \emph{precise} clock is new. We dedicate this paper to a comparison of rotation-mass-age predictions and a critical evaluation of the limiting precision of the rotation-mass-age relationship.  We quantify two effects that limit the intrinsic precision of rotation-age relationships: the large initial range of stellar rotation rates and surface differential rotation. We also investigate the sensitivity of the age errors to different theoretical approaches and compare with intermediate aged and old cluster systems.

\subsection{Rotation Data and Gyrochronology}\label{sec:introgyro}

The qualitative rotation-mass-age relationships from earlier decades are now being re-evaluated as practical tools for stellar population studies, and open cluster stars have once again become a central calibrating tool. Both large scale monitoring campaigns driven largely by planet transit searches and dedicated stellar rotation programs, such as the Monitor Project (\citealt{Irwin 2006,Irwin 2007a,Irwin 2007b,Irwin 2008a,Irwin 2008b} and \citealt{Meibom 2009,Meibom 2011a,Meibom 2011b}), have built on pioneering period searches in clusters (e.g.\ \citealt{Radick 1987,van Leeuwen 1987}.) Modern rotation-mass-age, or gyrochronology \citep{Barnes 2003}, relations \citep{Pace 2004,Barnes 2007,Mamajek 2008,Meibom 2009,Meibom 2011a} are based on the following major insights gleaned from open cluster studies.

Protostars are born with a wide range of rotation rates (e.g.\ NGC 6530, \citealt{Henderson 2012}; Orion Nebula Cluster, \citealt{Attridge 1992}; IC 348, \citealt{Herbst 2000}). However, by the time young stars arrive on the main sequence, the majority of them are rotating at similar, and slow, rotation rates (\citealt{Stauffer 1989} and \citealt{Hartman 2010} are nice examples for the Pleiades cluster.) One can therefore define a median relationship that is applicable to many solar analogs, even young ones. The median rotation period at fixed color increases with increased age. Additionally, the width of the rotation distribution also narrows with time; this process is fastest for solar analogs and takes progressively longer for lower mass stars. This convergence is seen in solar analogs by ages of order 500 -- 600 Myr, exemplified by the Hyades \citep{Radick 1987} and M37 \citep{Hartman 2009}. Published gyrochronology prescriptions take advantage of these basic features to define working relationships between rotation, color, and age using a semi-empirical approach. The median rotation of the bulk population is defined for young open clusters, tested in clusters with ages under 1 Gyr, and extrapolated to older stars using the solar calibration as a constraint. Gyrochronology relationships provide an empirical expression for rotation as a function of color with a time dependence based on \citet{Skumanich 1972}.

Gyrochronology is a promising concept, but there is real tension between our understanding of angular momentum evolution, cluster data, and a simple mapping of rotation onto a mass-dependent age. This is especially true for rotation-mass-age relationships in the context of young clusters, which we examine in a companion paper (Epstein, Pinsonneault, \& Terndrup 2013; Paper II.) These theoretical issues are important because the simple empirical fits described above make strong assumptions about the underlying physics, and these assumptions are not consistent with the results of detailed studies.  We therefore begin with a brief review of modern models of angular momentum evolution, and then use their properties as guidance for a more physically motivated approach to gyrochronology.

\subsection{A Standard Picture of Angular Momentum Loss in Low Mass Stars}\label{sec:introstdpic}

The theory of angular momentum evolution is complex. To predict rotation-age relationships, one must understand the origin of stellar rotation, angular momentum transport in stellar interiors, and angular momentum loss from magnetized solar-like winds. 

Open clusters and star forming regions have been crucial for developing the standard picture of angular momentum evolution in low mass stars. Large rotation datasets, and models capable of addressing them, became available in the late 1980s (\citealt{Stauffer 1989,Pinsonneault 1989}). Rotation measurements in star forming regions set the initial conditions for angular momentum evolution, while open clusters provide snapshots in time of the evolution of stellar rotation rates.

A standard working model of angular momentum evolution has emerged, with some important refinements on the simplest possible model. Protostars of the same mass are born with a range of rotation rates (e.g.\ \citealt{Edwards 1993,Bouvier 1986}; for more recent work see \citealt{Rebull 2004}, \citealt{Henderson 2012}). The simplest model of angular momentum evolution would begin with the distribution of rotation rates in protostars and evolve them forward assuming a scaled solar angular momentum loss rate $\mathrm{d}J/\mathrm{d}t \simeq \omega^{3}$ \citep{Kawaler 1988}, which reproduces the \citet{Skumanich 1972} law.  Early models of this form \citep{Pinsonneault 1990} were inconsistent with open cluster data, and this pointed to three major refinements of the picture.  

Star-disk interactions, grouped under the general label of disk-locking \citep{Koenigl 1991,Edwards 1993,Shu 1994}, are required to understand the mapping from protostars to young open cluster stars.  For example, \citet{Tinker 2002} showed that disks with a maximum lifetime between 3 and 8 Myr can reconcile the protostellar and ZAMS rotation distributions.

The survival of rapid rotation in young stars (e.g. \citealt{Alphenaar 1981,Meys 1982,van Leeuwen 1982,Soderblom 1983,Stauffer 1984,Stauffer 1987}) prohibits a simple scaling of the solar torque by the cube of the rotation rate; saturation in the loss rate is required \citep{Chaboyer 1995,Barnes 1996}. A similar behavior is observed in proxies of magnetic activity \citep{Noyes 1984,Stauffer 1994,Pizzolato 2003,Wright 2011} and expected in wind models \citep{MacGregor 1991}. Both the activity and spin down data require a mass dependent saturation threshold \citep{Krishnamurthi 1997,Sills 2000}, in the sense that lower mass stars saturate at lower rotation rates.

The final additional ingredient is the detection of core-envelope decoupling. Stars spin up as they contract onto the main sequence, where their structure stabilizes and the timescale for evolution increases. A magnetized wind will initially spin down the convective envelope; the radiative core will then respond over the internal angular momentum transport timescale. There will be a rapid initial drop in surface rotation if the transport timescale is long enough, followed by an epoch where the core spins down and the envelope does not (e.g.\ \citealt{Endal 1979}). Once the entire star is coupled it will spin down as a solid body, even if there is a differentially rotating internal profile \citep{Pinsonneault 1989}. The observational signature is that the slow rotator population in young open clusters spins down quickly at first, and then the spin down stalls. Numerous studies have confirmed this phenomenon \citep{Endal 1981,Pinsonneault 1990,MacGregor 1991conf,MacGregor 1991,Keppens 1995,Jianke 1993,Charbonneau 1993,Keppens 1995,Krishnamurthi 1998,Allain 1998,Irwin 2007b,Bouvier 2008,Denissenkov 2010,Spada 2011}, with inferred timescales in the range of tens to hundreds of Myr, depending on mass. Rapid rotators appear to always spin down as a solid body, suggesting that the transport timescale is a function of both rotation rate and mass (e.g.\ \citealt{Denissenkov 2010,Spada 2011}). The underlying mechanisms for all three of these effects are active research topics, but they are all physically well-motivated and supported by strong empirical evidence. A practical consequence is that there are documented problems with adopting a Skumanich-style spin down model in young stars.

\subsection{Our Approach} 

Fortunately, there is strong evidence that the physical picture is simpler for the older stars that will dominate a field sample. Protostar-accretion disk interactions are important on timescales of millions of years, while effective core-envelope coupling is established by an age of 500 Myr (e.g.\ \citealt{Denissenkov 2010}).

In this paper, we therefore focus on the expected behavior of old field stars, with an emphasis on stars older than 500 Myr. The distribution of initial conditions can be empirically specified by open cluster stars, and the assumption of solid body spin down at this age and beyond is well-justified. Furthermore, there are good constraints on angular momentum loss, permitting a well-posed forward modeling of angular momentum evolution.

Much recent work has focused on presenting gyrochronology recipes as a function of color. We compare these relationships with two angular momentum loss models and the extremely limited current data for old stars. New data, especially from the Kepler mission, will shed valuable light on the actual functional form of rotational evolution in old stars (a point that we will return to in the conclusions.) Here, we emphasize an under-appreciated aspect of the problem: \emph{how precise are rotational ages?}

This paper is organized as follows. Section \ref{sec:method} contains a detailed description of the initial conditions, stellar interior model, braking law, and assumptions necessary to model stellar angular momentum evolution. We present our results in \S\ref{sec:results}. We compare the median rotation-mass-age relationship predicted by our preferred angular momentum loss model with an alternate braking law (\S\ref{sec:timeEvolution}) and published gyrochronology relations (\S\ref{sec:compare2gyrochronology}). Sources of uncertainty in obtaining rotation-based ages are quantified in \S\ref{sec:uncertainties}. These include both fundamental limitations (e.g.\ arising from the initial scatter in rotation rates, \S\ref{sec:ageatfixedP}, and from surface differential rotation, \S\ref{sec:minPeriod}) and model-specific errors (\S\ref{sec:distwidth}).  We compare our models with open cluster rotation period data in \S\ref{sec:Intermediate-AgeCluster}.  Those theoretical predictions are tested against observational constraints from two binary star systems 61 Cyg and $\alpha$ Cen in \S\ref{sec:oldfieldstars}. 

Kepler and CoRoT are measuring asteroseismic ages for tens of thousands of old solar-like stars (e.g.\ \citealt{Mosser 2010,Gilliland 2010}). However, such space missions cover only a limited number of fields and targets. We therefore also examine how the prospects for rotation-mass-age relationships depend on the sensitivity of the survey. In \S\ref{sec:conclusions}, we demonstrate that ground-based studies can find young to intermediate aged populations (based on variability) for plausible precision ranges, while it is likely that the detection of older populations will require a degree of precision currently achievable only from space. We also compare the limiting precision of rotation-based ages with other age diagnostics: trigonometric parallaxes and asteroseismology.

\section{Method}\label{sec:method}

We investigate two observational effects that limit the precision of rotation-age relationships: the range of rotation rates at fixed age and the precision of rotation period measurements. This requires an angular momentum evolution model and a means of specifying the range of rotation rates. Our adopted initial conditions and treatment of differential rotation are described in \S\ref{sec:initial conditions}.  We provide the details of our stellar interiors and angular momentum evolution models in \S\ref{sec:interiors} and \S\ref{sec:torque}, respectively.

\subsection{Empirical Constraints}\label{sec:initial conditions}

We discuss the benefits of using the intermediate-aged cluster M37 to set the initial conditions for the angular momentum loss models in \S\ref{sec:M37}. We also introduce other intermediate-age clusters that can be used to constrain our angular momentum evolution models (\S\ref{sec:Intermediate-AgeClusterData}). We use observations of old field stars conducted by the Mount Wilson survey to determine the significance of differential rotation (\S\ref{sec:Old Field Star DR}).

\subsubsection{Initial Conditions: M37}\label{sec:M37}

There are three natural starting points for angular momentum evolution studies. The protostellar domain is the natural physical one, but forward evolution requires modeling of the interaction between protostars and disks. We could also begin with young open clusters, but it is clear from prior work that the surface rotation rates do not reflect the internal angular momentum content. We have dedicated a separate paper to studying this (different) problem  and the implications for rotation-age relationships in young stellar populations. By contrast, solar analogs in the 500 Myr and older regime can be treated as more strongly coupled, permitting a much less model dependent forward approach. We therefore bypass model-induced uncertainties in the initial rotation distribution by adopting an empirical rotation distribution drawn from measurements of stellar rotation in the intermediate-aged open cluster M37 (NGC 2099). The vast majority of field stars will be older than 500 Myr, which makes this simpler approach attractive for examining field star rotation studies.

We selected M37 because it is the cluster with the largest homogeneous set of measured rotation periods spanning a wide color range in the intermediate age range. Using an empirical distribution to set the initial conditions, however, does mean that the results are sensitive to the same biases and systematics present in the input cluster data. Fortunately, M37 has well understood biases, measures of completeness, and little field contamination. \citet{Hartman 2008,Hartman 2009} performed a deep, 20 night survey with the MMT of M37, reaching down to $r\sim23$ mag. They determined an age of $550\pm30$ Myr (with overshooting), metallicity [M/H]=$+0.045\pm0.044$, and distance modulus $(m-M)_V=11.572\pm0.13$. reaching down to $r\sim23$ mag. Searching for periodic variability using a multiharmonic AoV algorithm \citep{Schwarzenberg 1996} yielded a clean sample of 372 rotation period measurements. The sample is close to 100\% complete to $r=20$ mag (see Figure 9 in \citealt{Hartman 2009}), in the sense of detecting stars with modulation above 0.01 mag, although the color-period diagram may be biased toward brighter stars. This means that an increasing fraction of cluster members have undetectable variations as mass declines. Hartman et al.\ estimate $\sim20\%$ contamination of the cluster sample by background field stars.

\citet{Hartman 2008} obtained B, V, and I$_C$ band photometry for this cluster. For a given set of isochrones, different colors resulted in consistent mass estimates in regimes where the isochrones were not double-valued. For this paper, we select (V--I$_C$)$_0$ to estimate mass because it has a large dynamic range and its relationship to mass remains monotonic for cool stars. The (V--I$_C$) photometry has been corrected for reddening E(V--I$_C$)=0.355 \citep{Hartman 2008} and compared to the corresponding 550 Myr, [Fe/H$]=+0.045$ YREC isochrone. The implications of our choice of isochrone for the forward modeling are discussed in Section 3.

\subsubsection{Other Intermediate-Age Cluster Data}\label{sec:Intermediate-AgeClusterData}

We will test the common assumption that clusters form an evolutionary sequence by comparing the rotation distribution predicted by our angular momentum evolution model (anchored at M37) with observations in the Hyades, Praesepe, and NGC 6811. Rotation-based clocks are based on the premise that stars share a common underlying rotation distribution at birth and their subsequent angular momentum evolution is governed by the same physics. In this paradigm, each cluster is a realization of the underlying stellar rotation distribution at a given age. If cosmic scatter in the initial distribution of rotation rates is small, the predicted rotation distribution at a future time will be insensitive to the choice of initial conditions. Here, we describe the cluster data used to conduct this test; see \S\ref{sec:Intermediate-AgeCluster} for results.

\paragraph{Hyades and Praesepe}

The Hyades and Praesepe are the oldest clusters for which rotation period measurements have been measured with ground-based telescopes. The Hyades was one of the first open clusters targeted by rotation period surveys \citep{Radick 1987,Prosser 1995}. However, the large angular extent of the cluster makes it difficult to monitor large numbers of cluster members. Recently, \citet{Delorme 2011} doubled the number of cluster stars with measured rotation periods by using lightcurves from the Wide-Angle Search for Planets (SuperWASP) survey.  \citeauthor{Delorme 2011} report rotation periods for 62 members of Hyades and 52 members of Praesepe, where the cluster membership probability was determined by comparing proper motion and apparent magnitude with expectations. Praesepe has been a beehive of activity with three rotation period studies published within one year \citep{Delorme 2011,Scholz 2011,Agueros 2011}\footnote{We do not include the \citet{Scholz 2007} sample of rotation in Praesepe because of its small size (5 stars) and lack of position and photometric information.}. Together, this provides a sample of 87 Hyades members and 116 Praesepe members.

B- and V-band photometry is not available for all Hyades and Praesepe members, so we instead utilize 2MASS photometry for these two clusters to bin the sample by color and compare to simulated rotation distributions. We adopt E(B--V)=0 for Hyades and Praesepe (\citealt{Crawford 1966,Crawford 1969}, respectively, following \citealt{Pinsonneault 1998b}.)

Because the Hyades and Praesepe share similar ages, metallicities, and kinematics (e.g.\ \citealt{Eggen 1992}), it has been suggested that they formed from the same molecular cloud. \citet{Perryman 1998} found an age of $625\pm50$ Myr age for the Hyades using isochrones with with convective overshooting.  \citet{Claver 2001} favor an age for Praesepe close to that of the Hyades, between 500 and 700 Myr. \citet{Salaris 2004} places both the Hyades and Praesepe at $700\pm100$ Myr.  Additional age estimates for Praesepe include 650 Myr (e.g. \citealt{Scholz 2007}) and $757\pm36$ Myr \citep{Gaspar 2009}.

\paragraph{NGC 6811}

The amplitude of photometric modulations due to spots declines with time and space-based missions are necessary to measure rotation periods among older populations of stars. The Kepler Cluster Study released its first results for 71 members of the open cluster NGC 6811 using the first four quarters of Kepler data (Q1-Q4; \citealt{Meibom 2011b}.) \citeauthor{Meibom 2011b} transformed the SDSS g--r band photometry from the Kepler Input Catalog \citep{Brown 2011} to $(\mathrm{B - V})_0$ using equations from \citet{Jester 2005} and assuming E(B--V)=0.1. We adopt the measured rotation periods and photometry provided in \citet{Meibom 2011b}.

The age of NGC 6811 is not well known. \citet{Lindoff 1972} obtained the first photometry of this cluster and derived an age of 500 Myr. The cluster's age estimate was revised upwards to $700\pm100$  Myr \citep{Barkhatova 1978,Glushkova 1999}. A decade later, \citet{Luo 2009} measured an even younger age of $\log(t)=8.76\pm0.09$ (or $575 ^{+130}_{-110}$ Myr) with CCD observations of 600 stars in the cluster field. \citet{Meibom 2011b} adopts an age of 1 Gyr for NGC 6811 based on an unpublished data from the WIYN Open Cluster Study \citep{Matson 2005}. Additionally, \citet{Meibom 2011b} use the procedure developed by \citep{Janes 2011} to estimate an age of $1.1\pm0.2$ Gyr from the color of the main-sequence turnoff and the red clump. Unfortunately, details of the age determination process are not supplied. One possible cause for this $\sim500$ Myr dispute over the age of the cluster is significant contamination of the CMD by background field stars. \citet{Sanders 1971} provides cluster membership probabilities based on relative proper motions, but \citet{Meibom 2011b} are the first to perform a radial velocity test for membership in this cluster.

\subsubsection{Surface Differential Rotation and Observed Period Errors}\label{sec:Old Field Star DR}

We will examine how latitudinal differential rotation affects the precision of rotation period measurements and, by extension, rotation-based ages. If the Sun's differential rotation is characteristic of solar type stars, then the latitude of a star spot could significantly impact the measured rotation period. Differential rotation in field stars is best characterized by the Mount Wilson survey, with observations of stars spanning years to over a decade (e.g.\ \citealt{Donahue 1996}, see \S\ref{sec:oldfieldstars}).
Ideally, this would provide the uncertainty in rotation period measurements due to differential rotation. However, because \citet{Donahue 1996} only report the minimum, maximum, and mean observed rotation period, we cannot know the shape of the period distribution.

We therefore performed a simulation to determine the relationship between the range of observed rotation rates ($\Delta \mathrm{P}$) and the uncertainty in the mean period ($\sigma$). We randomly drew periods from the underlying distribution, which for simplicity we modeled as a Gaussian\footnote{In actuality, this distribution is probably a convolution of many different effects, including the range of latitudes sampled by spots at different points in the stellar cycle (e.g. the butterfly diagram), visibility effects, and variation in rotation rates at fixed latitude.}.  We calculated $\Delta \mathrm{P}$ as a function of the number observing seasons and compared it to the standard deviation of the underlying Gaussian. As the number of observations of the star increases, so too does the likelihood that you catch the star in one of the lower probability states (e.g.\ with a polar star spot), leading to a larger $\Delta \mathrm{P}$. After performing 1000 trials, we computed the mean $\left\langle \mathrm{P} \right\rangle-\sigma$ relationship.

We then used this tabulated relationship to determine an effective period uncertainty for the \citet{Donahue 1996} sample. The fractional uncertainty ranges from $\sigma_P/\mathrm{P}=2\%$ to 16\%, with an average of 5\%. We find weak evidence for a linear trend, where the fractional uncertainty increases with mean period. For slowly rotating stars ($\mathrm{P}>10$ days), where differential rotation is a larger effect, the mean fractional uncertainty is 7\%.

We model the impact of differential rotation as setting a lower limit on the range of rotation periods. Because we use the interquartile range rather than a standard deviation to quantify the range of rotation periods, we correct for the fact that the interquartile range falls within $\pm0.6745\sigma$. For illustrative purposes, we adopt a minimum fractional uncertainty, such that with the correction, latitudinal differential rotation sets a lower limit on the precision of rotation period measurements of $ 2 \times 0.6745 \times \sigma_P/P=10 \%$. This is just slightly larger than would be found using the mean fractional uncertainty, but well within the range found in the simulation of \citet{Donahue 1996} data described above. 

\subsection{Interiors}\label{sec:interiors}

We anchor our angular momentum study in theoretical models of stellar structure calculated using the YREC stellar interior models. We compute solar metallicity models (X=0.7190, Z=0.0165) with no diffusion. These provide a prior on a star's He abundance, radius, moment of inertia, convective overturn timescale ($\tau_{CZ}$), and luminosity as a function of stellar mass and age.

The standard model of angular momentum loss described in \S\ref{sec:introstdpic} is only valid for stars in a particular mass range. Because stars above spectral type F8 show no obvious sign of a rotation-age relationship \citep{Wolff 1986}, this sets an upper mass limit on where rotation-based ages are well motivated. For cooler stars, the saturation threshold appears to scale with the ratio of the rotation period to the overturn timescale, called the Rossby number ($R_0=\mathrm{P}_\mathrm{rot}/\tau_{CZ}$). However, there is evidence for a departure from a Rossby-scaled saturation threshold in the low mass regime (\citealt{Sills 2000}; see \S\ref{sec:kawaler}). Furthermore, quantities like $\tau_{CZ}$ become ill-defined for fully convective stars. The old open clusters also have sparse data at the low mass end. We therefore restrict our calculations to the mass range of $0.4\ \mathrm{M}_\odot \leq M \leq 1.2\ \mathrm{M}_\odot$, sampled in equally spaced 0.05 M$_\odot$ intervals. Stars in M37 (used to define the initial rotation distribution, \S\ref{sec:M37}) are excluded if their mass lies outside of this restricted range where our angular momentum loss model is valid.

The empirically determined stellar masses and rotation rates from M37 set the initial conditions (\S\ref{sec:M37}). A prescription for angular momentum loss (\S\ref{sec:torque}) is used to simulate M37's rotational future. We calculate a star's angular momentum at each point in time on a logarithmically-spaced grid of ages. At fixed age, we perform a polynomial interpolation in logarithmic rotation period on a grid of models to the mass of the star.

We set the birth line by requiring that the deuterium abundance declines to a level of 0.1\% below the initial value, resulting in a slight offset from the \citet{Sills 2000} models. We stop our models if their age exceeds either a stars' lifetime on the main sequence or the age of the Galactic disk, assumed to be 10 Gyr. The main sequence turnoff is defined as the point when the core hydrogen abundance drops below $X_C=10^{-3}$.

In accordance with standard mixing length theory, we define $\tau_{CZ}$ locally as $\tau_{CZ}=H_p/v_{conv}$ evaluated one pressure scale height above the base of the surface convection zone. In the YREC interior calculations, there is numerical jitter in $\tau_{CZ}$ at the $\lesssim1$\% level over a stars's main sequence lifetime. We therefore average $\tau_{CZ}$ over the five closest points in time to smooth over these numerical perturbations in the depth of the convection zone.

\subsection{Angular Momentum Evolution}\label{sec:torque}

We adopt solid body rotation in radiative and convective regions for our angular momentum transport treatment.  The latitude-averaged solar convection zone rotation is consistent with this approximation.  Core and envelope coupling timescales are short in comparison to the lifetime of G, K, and M dwarfs discussed here (e.g.\ \citealt{Denissenkov 2010,Spada 2011}). Because we are studying the rotation of field stars, assuming that the coupling timescale for angular momentum transport between the core and envelope is short (i.e. solid body rotation is enforced at all times) is a reasonable approximation for the majority of the age range of interest.

\subsubsection{Kawaler Wind Law}\label{sec:kawaler}

To simulate stellar angular momentum loss, we adopt as our preferred model a modified Kawaler wind law \citep{Chaboyer 1995,Sills 2000}. \citet{Kawaler 1988} was able to recover the \citet{Skumanich 1972} law ($\omega \propto t^{-1/2}$) for his favored choice of the magnetic field geometry. This yields a wind law of the form $\mathrm{d}J/\mathrm{d}t=-K \omega^3 R^{1/2}M^{-1/2}$, where the normalizing constant $K$ parametrizes our ignorance about winds and magnetic field generation in stars. This produces a nearly mass-independent angular momentum loss rate. The strongly mass-dependent rotation data, especially at at $(B-V)<0.4$ and $(B-V) > 1.4$, led \citet{Barnes Kim 2010} to question the Kawaler loss law (see their Figure 6). The former likely occurs because the physics changes for high mass stars with a very thin surface convection zone, resulting in a different mass-dependence for stars above the Kraft break. The Kawaler angular momentum loss law is best suited to operate in the region below the Kraft break and above the fully convective limit, where stars may develop a different type of dynamo.

As initially conceived, this law does not account for magnetic saturation-- a necessary element in the standard theory of angular momentum loss (\S\ref{sec:introstdpic}). We follow \citet{Krishnamurthi 1997} and \citet{Sills 2000} in modifying the original Kawaler loss law to include a prescription for how the torque changes when a star's magnetic field strength saturates at a critical rotation rate $\omega_{crit}$. Because the \citet{Reiners 2012} wind law (discussed in the next section \S\ref{sec:reinerslaw}) treats magnetic saturation differently, we supply a brief discussion of the empirical motivation and prior theoretical work that produced the functional form of the saturation threshold adopted here.

Observational studies of rotation and magnetic activity indicators support a Rossby-scaled magnetic saturation threshold. Both chromospheric Ca \textsc{II} H and K and coronal X-ray emission display a break in their relationship with Rossby number \citep{Noyes 1984,Patten 1996}. In fact, activity correlates more tightly with Rossby number than rotation period \citep{Radick 1987}. Because measures of stellar activity serve as a proxy for magnetic field strength, which in turn powers the wind, the rate of angular momentum loss is also thought to plateau. Consequently, \citet{Krishnamurthi 1997} proposed the addition of a Rossby scaled saturation threshold for stars more massive than 0.5 M$_\odot$, resulting in the following modified Kawaler wind law:

\begin{equation}\label{eq:kawaler}
\frac{dJ}{dt}=
\begin{cases}
-K \omega_{crit}^2 \omega(\frac{R}{R_\odot})^{\frac{1}{2}}(\frac{M}{M\odot})^{-\frac{1}{2}} & \mathrm{for}\ \omega>\omega_{crit}\\
-K \omega^3(\frac{R}{R_\odot})^{\frac{1}{2}}(\frac{M}{M\odot})^{-\frac{1}{2}} & \mathrm{for}\ \omega\leq\omega_{crit}
\end{cases}
\end{equation}
where
\begin{equation}\label{eq:omegacrit}
\omega_{crit}=\omega_{crit,\odot}\frac{\tau_{CZ,\odot}}{\tau_{CZ}}.
\end{equation}
From the solar mass YREC model, $\tau_{CZ,\odot} = 1.01\times10^{6}$ s. Because lower mass stars have deeper convection zones, $\tau_{CZ}$ contains the mass dependence in the saturation threshold.

X-ray measurements confine the saturation threshold $\omega_{crit}$ to the range 5 to 20 $\omega_\odot$ \citep{Patten 1996}. \citet{Sills 2000} tested the Rossby scaling by comparing theoretical models to rotation distributions of IC 2602, IC 2391, $\alpha$ Per, the Pleiades, and Hyades. They found that the \citet{Krishnamurthi 1997}\ scaling ($\omega_{crit,\odot}=10\ \omega_{\odot}=2.86\times10^{-5}\ \mathrm{s}^{-1}$ for solid body models) works for stars with masses between $0.6\ \mathrm{M}_\odot < M < 1.1\ \mathrm{M}_\odot$. However, the Rossby scaling breaks down for lower mass stars; \citet{Sills 2000}\ argue that the saturation threshold drops more steeply than a Rossby scaling for stars with 0.1 M$_\odot < \mathrm{M} < 0.5\ \mathrm{M}_\odot$. They then used the clusters to empirically measure the mass dependence of $\omega_{crit}$ for stars in this mass range. For solid body models, they find that $\omega_{crit,\odot}=7$, 6.2, 5.1, 3.6, and 1.7 $\omega_\odot$ at 0.5, 0.4, 0.3, 0.2, and 0.1 M$_\odot$. We interpolate on the \citet{Sills 2000}\ calibration for stars $\mathrm{M} \leq 0.5$ M$_\odot$ and adopt the \citet{Krishnamurthi 1997}\ calibration for M $> 0.5$ M$_\odot$.

Given this definition of $\omega_{crit}$, Eq.\ \ref{eq:kawaler} is calibrated such that a $1\ \mathrm{M}_\odot$ star achieves the solar rotation rate $\omega_\odot=2.86 \times 10^{-6}$ rad s$^{-1}$ (the Sun's equatorial rotation period is $\mathrm{P}_\odot = 25.4$ days) at the solar age of 4.5 Gyr starting with an initial period of 10 days and no disk locking. This results in $K_\odot=3.15\times 10^{47}$ g cm$^2$ s. The choice of initial rotation period is plausible; it is similar to the average T Tauri rotation period of 9.45 days measured by \citet{Choi 1996} in the ONC. However, the predicted rotation period at the age of the Sun, and therefore the value of the normalization constant, is highly insensitive to the choice of initial rotation period. This is a characteristic feature of angular momentum laws of with a strong dependence on rotation rate, namely, that once stars enter the unsaturated regime ($\omega \leq \omega_\mathrm{crit}$), they rapidly forget their initial conditions. For example, if we instead calibrated the solar rotation rate with an initial period of 10 days and a 5 Myr disk locking time, this would result in a slightly different normalization constant $K_\odot = 2.97\times 10^{47}$ g cm$^2$ s.

\subsubsection{\citeauthor{Reiners 2012} Wind Law}\label{sec:reinerslaw}

\begin{figure}
\begin{center}
\includegraphics[width=3.4in]{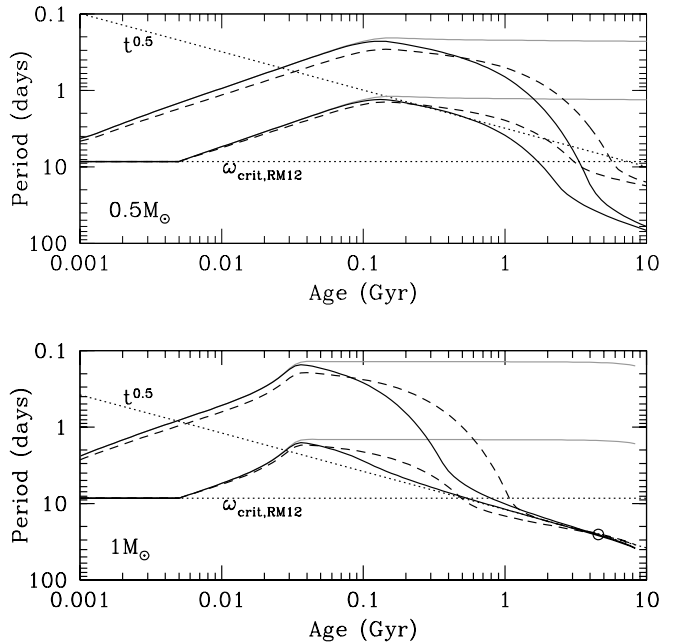}
\end{center}
\caption{Rotation period evolution according to the modified Kawaler (Eq.\ \ref{eq:kawaler}; solid) and \citeauthor{Reiners 2012} (Eq.\ \ref{eq:reiners}; dashed) braking laws for a 0.5 M$_\odot$ (top) and 1 M$_\odot$ (bottom) star. All models assume an initial rotation period of 8.5 days. Evolutionary tracks are shown for scenarios with 0 and 5 Myr of disk-locking. The \citeauthor{Reiners 2012} law has been re-calibrated such that it reproduces the Sun with the YREC interior models. The diagonal dotted line gives the solar calibrated Skumanich $P\sim t^{0.5}$ relation, calibrated to the Kawaler law at 550 Myr with 5 Myr disk-locking (top) and to the Sun (bottom). The horizontal dotted line shows the mass- and time-independent saturation threshold adopted by \citet{Reiners 2012}. Grey solid lines show the case of no braking for comparison. The rotation axis is reversed in this and future figures so that as stars spin down, they move down the page toward longer periods.} \label{fig:reiners}
\end{figure}

\citet{Reiners 2012} present a new braking law that pivots on two major deviations from the modified Kawaler law. First, \citeauthor{Reiners 2012} argue that rotation is related to the magnetic field strength ($B_0\propto \omega^a$) rather than the surface magnetic flux ($B_0 R^2\propto \omega^a$) as assumed in \citet{Kawaler 1988}. This factor of $R^2$ difference has powerful ramifications when propagated to the angular momentum loss law. Second, \citeauthor{Reiners 2012} assume that the saturation threshold $\omega_{crit}$ does not depend on mass. This differs from the Rossby-scaled saturation threshold used in \S\ref{sec:kawaler}; in this new law, the saturation threshold is insensitive to structural differences in stellar interiors. The reader should refer to \citet{Reiners 2012} to see the specifics of their derivation, but the resulting angular momentum loss law (their Eq.\ 5) is reproduced below:
\begin{equation}\label{eq:reiners}
\frac{dJ}{dt}=
\begin{cases}
-\ \mathcal{C}\ \omega \left(\frac{R^{16}}{M^2}\right)^{1/3} &{\rm for}\ \omega \ge \omega_{\rm crit} \\
-\ \mathcal{C} \left(\frac{\omega}{\omega_{\rm crit}}\right)^{4}\omega\left(\frac{R^{16}}{M^2}\right)^{1/3} &{\rm for}\ \omega < \omega_{\rm crit}
\end{cases}
\end{equation}
where $\mathcal{C}$ is a constant relating to the magnetic field strength and rate of mass loss, $R$ is stellar radius and $M$ is stellar mass in cgs units. It is important to note that \citeauthor{Reiners 2012} use the saturation threshold in the opposite sense of the modified Kawaler law. In the saturated regime ($\omega \geq \omega_\mathrm{crit}$), the rate of spin down is set by a maximum magnetic field strength (B$_\mathrm{crit}$) for all stars. The saturation threshold $\omega_{crit}$ only affects the torque in the \emph{unsaturated} regime ($\omega < \omega_{\rm crit}$). This means that changing the saturation threshold affects the late-time behavior of stars under the \citet{Reiners 2012} prescription. This is not true for a modified Kawaler law, where the effect of $\omega_\mathrm{crit}$ is only felt in the saturated regime and erased as stars forget their initial conditions.

To calibrate this wind law, \citeauthor{Reiners 2012} match the solar rotation rate at the solar age, starting with an initial rotation rate of 8.5 days. The presence of a disk is assumed to hold the rotation rate constant for 5 Myr. This yields their best fit combination of parameters [$\omega_{\rm crit}$,
$\mathcal{C}$] = [8.56$\times$10$^{-6}$ s$^{-1}$, 2.66$\times$10$^{3}$
(g$^5/$cm$^{10}$s$^{3}$)$^{1/3}$]. We recalibrate the constant $\mathcal{C}$ to reproduce the Sun with our models and a 5 Myr disk-locking lifetime, finding a different value of $\mathcal{C}$ = 3.16$\times$10$^2$ (g$^5/$cm$^{10}$s$^{3}$)$^{1/3}$.

Figure \ref{fig:reiners} compares angular momentum evolution under both a modified Kawaler and a \citeauthor{Reiners 2012} wind law with and without a disk. The largest differences appear among low mass stars. As discussed in \S\ref{sec:kawaler}, \citet{Sills 2000} find that the Rossby scaling is inadequate for stars $<0.6$ M$_\odot$. The \citeauthor{Reiners 2012} law naturally preserves rapid rotation of low mass stars. For solar mass stars, the Kawaler wind law asymptotes to the Skumanich relationship, becoming nearly indistinguishable by 500 Myr in the disk-locked case. In contrast, the \citeauthor{Reiners 2012} wind law predicts a slower spin down than Kawaler does for the first few Myr, followed by an interval of rapid angular momentum loss between $\sim200 - 700$ Myr, before leveling off and approaching a rate slightly shallower than Skumanich later in life.

\section{Results}\label{sec:results}

We begin with the rotation distribution of M37 and run the angular momentum evolution models (described in \S\ref{sec:torque}) forward in time  (\S\ref{sec:timeEvolution}) and compare the resulting median relationship with previous rotation-age relationships (\S\ref{sec:compare2gyrochronology}). We then establish our error budget (\S\ref{sec:uncertainties}). The dominant source of uncertainty for low mass stars proves to be the intrinsic scatter in the initial rotation distribution. Because these stars require many Gyr to converge to a unique rotation-mass-age relationship, this induces large uncertainties in rotation-based ages. Among old stars, the range of surface rotation rates observable due to latitudinal differential rotation limits the precision. We verify that our choice of initial conditions and loss laws does reproduce the rotation distribution observed in other intermediate-aged clusters (\S\ref{sec:Intermediate-AgeCluster}). Although observational constraints on the rotation-mass-age relationship are rare at late times, two binary star systems, 61 Cyg and $\alpha$ Cen, possess independent age measurements. We therefore test the rotation-based age predictions from our models and published gyrochronology relationships against these binaries in \S\ref{sec:oldfieldstars}

\subsection{Median Rotation-Mass-Age Relationships}\label{sec:MedianRelationships}

Young clusters display a large spread in stellar rotation rates across all masses. This wide range converges as stars age, but the timescale for this process can be long, especially in lower mass dwarfs. We begin by modeling the time evolution of M37's rotation distribution and investigating how long it takes for rotation rates to converge. The timescale for convergence is important because the median rotation-mass-age relationship only becomes a good chronometer after stars have forgotten their initial conditions.

\subsubsection{Time Evolution of M37's Rotation Distribution}\label{sec:timeEvolution}

\begin{figure}
\plotone{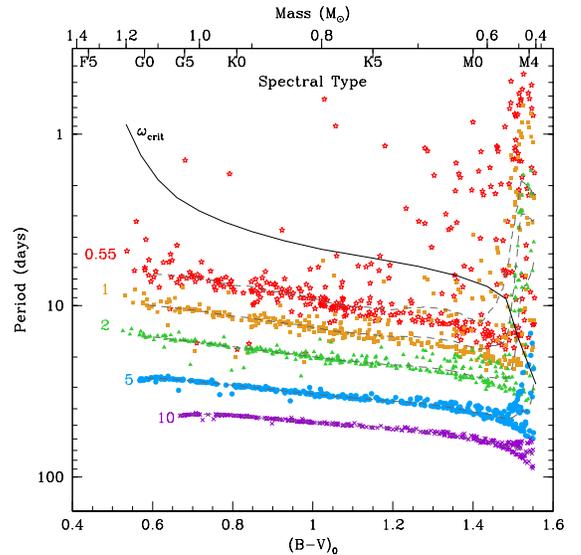}
\caption{M37's color-period distribution \citep{Hartman 2009} evolved forward in time from 0.55 Gyr (red stars) to 1 Gyr (gold squares), 2 Gyr (green triangles), 5 Gyr (blue circles), and 10 Gyr (purple crosses). The corresponding ages (in Gyr) are indicated to the left of the mass-rotation sequence. These synthetic rotation distributions are created by evolving the M37 rotation distribution forward in time with the modified Kawaler spin down model described in \S\ref{sec:kawaler}. The black curve shows the critical rotation rate ($\omega_{crit}$ at 550 Myr) above which stars fall in the saturated regime. The dashed curves show the median rotation period at each age. The associate mass and spectral type are given along the top axis. Masses are taken from a 1 Gyr, [Fe/H]$_\mathrm{M37}$=+0.045 YREC isochrone. Convergence occurs later for redder (lower mass) stars.}\label{fig:M37movie}
\end{figure}

\begin{figure}
\begin{center}
\plotone{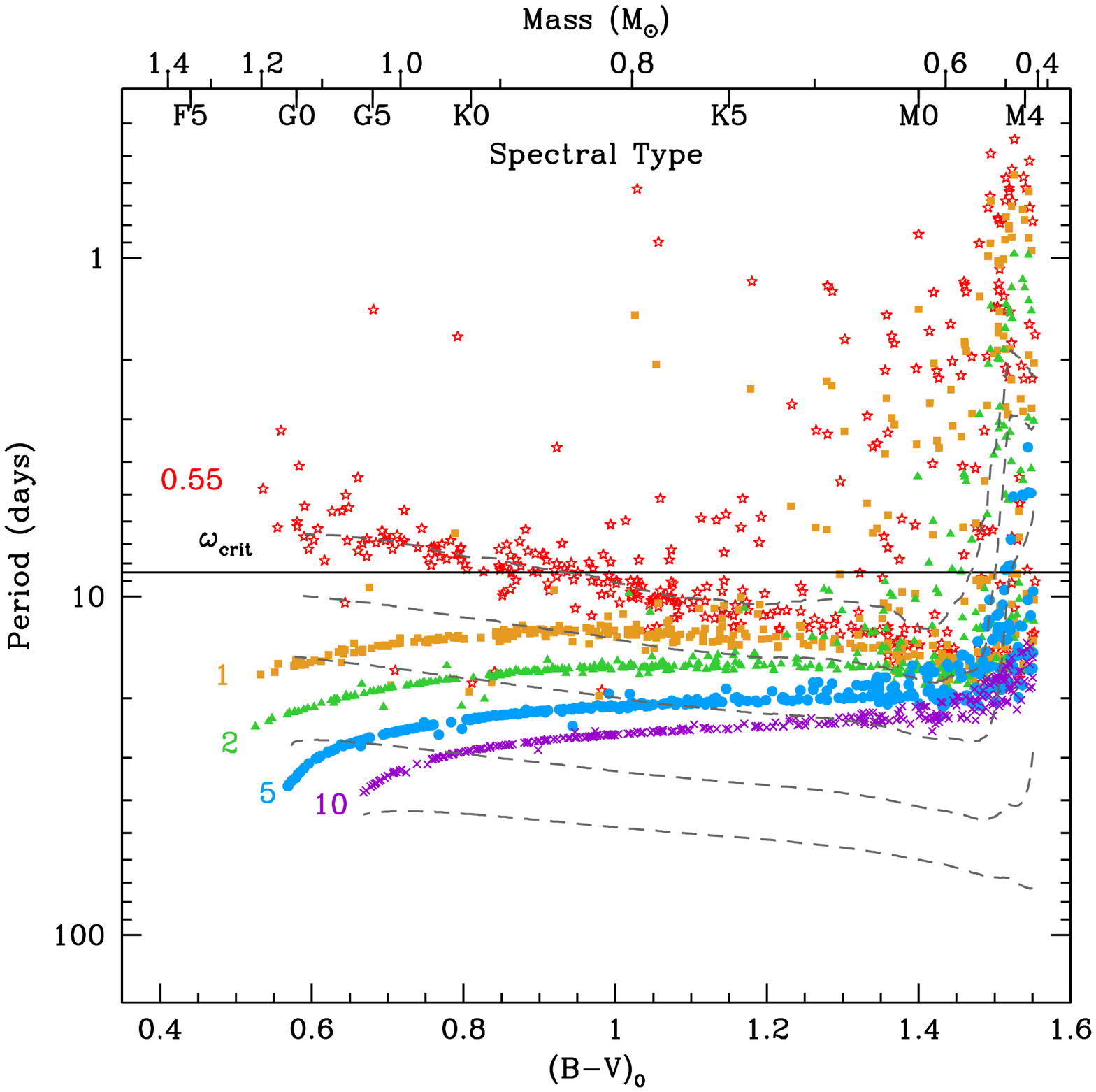}
\end{center}
\caption{In analogy to Figure \ref{fig:M37movie}, M37's color-period distribution \citep{Hartman 2009} evolved forward in time from 0.55 Gyr (red stars) to 10 Gyr (purple crosses) according to the \citet{Reiners 2012} braking law described in \S\ref{sec:reinerslaw}. The corresponding ages (in Gyr) are indicated to the left of the mass-rotation sequence. The black line shows the constant critical rotation rate ($\omega_{crit}$=8.56$\times$10$^{-6}$ s$^{-1}$). The dashed lines indicate the corresponding medians from the Kawaler wind law (Figure \ref{fig:M37movie}) for comparison. The associate mass and spectral type are given along the top axis. Masses are taken from a 1 Gyr, [Fe/H]$_\mathrm{M37}$=+0.045 YREC isochrone.} \label{fig:movieReiners}
\end{figure}

The theoretical framework in \S\ref{sec:method} allows us to predict the rotation of a star at any future time, given its mass and initial period. We use solid body models to evolve M37's rotation distribution from 550 Myr forward in time. We explore two different prescriptions for the torque, starting with a modified Kawaler wind law (\S\ref{sec:kawaler}). Figure \ref{fig:M37movie} shows M37's initial rotation distribution at 550 Myr and also how that distribution would evolve under the influence of a modified Kawaler wind law. For visual clarity, we show the projected distribution at representative ages 1, 2, 5, and 10 Gyr. Even in the initial cluster distribution at 550 Myr, bluer stars have begun to form a mass-rotation sequence. Because of the rotation-dependent rate of angular momentum loss (Eq. \ref{eq:kawaler}), this sequence becomes tighter over time and redder stars converge to join it.

The vertical displacement of the coeval distributions reflects the fact that stars spin more slowly with time. The reddest stars in the cluster exhibit a large dispersion in rotation rate at young ages, but this range also shrinks over time. Nonetheless, a star with spectral type M4 and a rotation period of 20 days easily falls within the overlapping range of the 0.55, 1, and 2 Gyr distributions, but could also be among the most rapidly rotating members of even older populations. This inherent ambiguity represents a fundamental uncertainty in rotation-based ages and will be discussed in depth in \S\ref{sec:ageatfixedP}.

The median mass-rotation curves at a series of ages may be inverted to provide rotation-based ages when the scatter about the median is small. However, we note that rotation rates still encode age information, even in situations when a star's mass and rotation rate does not correspond to a unique age. If the form of the distribution was known, and the rate of change calibrated, one could in principle obtain statistical information about ages even when the rotation-age solution is not unique. A probability distribution for the age of a star with a given mass and rotation period could be constructed by weighting these iso-age rotation distributions by the rate of star formation and the rate of angular momentum evolution. Given the uncertainties associated with the rates necessary to this calculation, we instead focus on the median rotation-mass-age relationship here and use the range of rotation to quantify the uncertainty in rotation based ages in \S\ref{sec:uncertainties}.

For comparison, we also explore the the effect of using a different angular momentum loss law. We begin with the same M37 rotation distribution at 550 Myr, but this time we use the \citet{Reiners 2012} wind law (\S\ref{sec:reinerslaw}) to model angular momentum evolution. Figure \ref{fig:movieReiners} shows how the shape of the M37 distribution changes under the \citeauthor{Reiners 2012} wind law compared to the modified Kawaler wind law (Figure \ref{fig:M37movie}). Conspicuously, the slope of the mass-rotation relationship appears to flip between 550 Myr and 1 Gyr. This arises from a combination of negligible spin down for low mass stars and an interval of rapid spin down for solar-type stars as noted in Figure \ref{fig:reiners}. After 1 Gyr, the rate of angular momentum loss predicted by the \citeauthor{Reiners 2012} law slows for stars of all masses, maintaining the inverted slope of the mass-rotation sequence established earlier.

The dashed lines in Figure \ref{fig:movieReiners} reproduce the median mass-rotation sequences from the Kawaler law. The difference between the \citeauthor{Reiners 2012} and Kawaler sequences is striking. In addition to having slopes of a different sign, the \citeauthor{Reiners 2012} law produces a smaller torque than the Kawaler law. To wit, the Kawaler law spins M $<0.9$ M$_\odot$ stars down to rotation periods slower than they achieve in 10 Gyr with the \citeauthor{Reiners 2012} law-- in half the time! Because of the compressed dynamic range of rotation rates in the \citeauthor{Reiners 2012} law compared to Kawaler, the overlap between iso-age populations is more extreme for low mass stars. The reddest stars have such weak predicted spin down with a \citeauthor{Reiners 2012} law that the 550 Myr and 10 Gyr distributions overlap. In this regime where spin down failed to separate iso-age populations, the overlap creates ambiguous rotation-based age predictions covering a wide range. To extract more information, one must rely on statistical ages.

\begin{figure}
\plotone{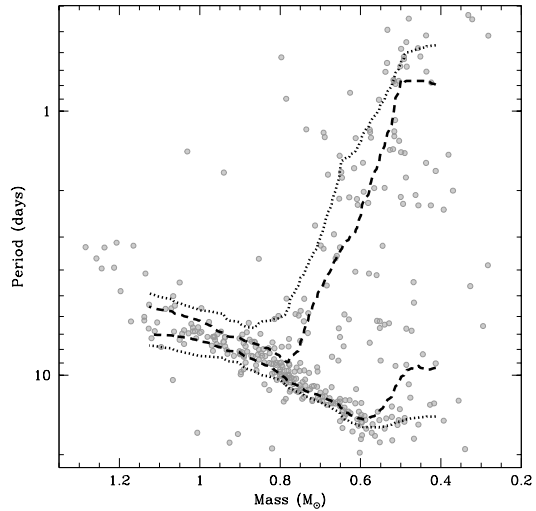}
\caption{Mass-period distribution for M37 sample \citet{Hartman 2009}. The dashed band marks the smoothed 25$^\mathrm{th}$ and 75$^\mathrm{th}$ period percentiles. The dotted band depicts the smoothed 10$^\mathrm{th}$ and 90$^\mathrm{th}$ period percentiles for comparison purposes.}\label{fig:M37percentile}
\end{figure}

Now that we have two predictions for how M37's rotation distribution evolves in time, we need to characterize their median trend and range. To quantify the width of the rotation distribution, we adopt the interquartile range in period. We calculate the 25$^\mathrm{th}$ and 75$^\mathrm{th}$ percentile in a moving window of 25 stars per mass bin and perform a smoothed boxcar average with a width of 50 stars to flatten out any star-to-star jitter in the calculated percentiles. The middle 50\% of stars in the observed M37 cluster mass-rotation distribution lie within the interquartile band shown in Figure \ref{fig:M37percentile} and we use this to characterize the spread between fast and slow rotators throughout most of the paper. The median is calculated with the same bins and smoothing as for the interquartile region.

For comparison, we also consider the 10$^\mathrm{th}$ and 90$^\mathrm{th}$ percentile rotation periods because they set much broader bounds than the interquartile region, but are not overly sensitive to outliers in the rotation distribution. We calculate the 10$^\mathrm{th}$ and 90$^\mathrm{th}$ percentile in a moving window of 25 stars per mass bin and perform a smoothed boxcar average with a width of 80 stars. The middle 80\% of stars in the observed M37 cluster mass-rotation distribution fall within the light gray band shown in Figures \ref{fig:M37percentile}. The effect of choosing this alternative metric for the width of the rotation distribution is discussed in \S\ref{sec:distwidth}.

The percentile ranges in Figure \ref{fig:M37percentile} clearly illustrate that higher mass stars have converged to a narrow mass-rotation sequence by the age of M37. Toward lower masses, the distribution rapidly fans out to span over a decade in period. This sharp transition may be used to define the point of convergence for which more massive stars have forgotten their initial rotation rates and converged to a tight sequence. We choose to define the point of convergence as the place where the width of the distribution is comparable (within a factor of 2) to the measured period uncertainties. As we shall see later, surface differential rotation limits the precision of rotation period measurements. In \S\ref{sec:minPeriod}, we adopt a minimum uncertainty of 10\% to reflect the range of surface rotation rates at different latitudes. Thus, we define our criterion for convergence to be when the interquartile region is smaller than 20\% of the median rotation period. This also qualitatively reproduces the by-eye location of the abrupt widening of the distribution.

By this metric, a converged sequence is established down to (B--V)$_0=$1.08 (M=0.78 M$_\odot$) at the starting age of 0.55 Gyr. For the Kawaler wind law, the mass-rotation sequence is converged to (B--V)$_0=$1.17 (0.74 M$_\odot$) by 1 Gyr, to (B--V)$_0=$1.41 (0.62 M$_\odot$) by 2 Gyr, to (B--V)$_0=$1.50 (0.52 M$_\odot$) by 5 Gyr, and to (B--V)$_0=$1.54 (0.46 M$_\odot$) by 10 Gyr. By the same rule, the \citeauthor{Reiners 2012} distributions are converged to (B--V)$_0=$1.19 (0.73 M$_\odot$) by 1 Gyr, to (B--V)$_0=$1.33 (0.67 M$_\odot$) by 2 Gyr, to (B--V)$_0=$1.46 (0.58 M$_\odot$) by 5 Gyr, and to (B--V)$_0=$1.54 (0.45 M$_\odot$) by 10 Gyr. It is interesting that the location of the point of convergence is robust to within 0.06 M$_\odot$ even though these two models treat the saturation threshold differently. These points define the red border of the regime where the distribution of rotation rates has tightened enough that the median rotation-mass-age relationship can provide good constraints on age under one of these wind laws.

\subsubsection{Comparison to Published Gyrochronology Relationships}\label{sec:compare2gyrochronology}

In the previous section, we quantified both the median rotation-mass-age relationship and range of rotation rates for two different angular momentum loss laws. Here, we reconcile those trends with the literature. Published gyrochronology relationships ignore rapidly rotating stars and focus on describing the shape and time-evolution of the sequence of slowly-rotating stars in open cluster samples (for further discussion, see \S\ref{sec:introgyro} and references therein). For the purposes of comparison, we concentrate on the median trend and postpone further discussion of the width of the rotation distribution until \S\ref{sec:uncertainties}.

\begin{deluxetable*}{lllll}
\tablewidth{0pc}
\tablecaption{Parameters in Gyrochronology Models}
\tablehead{\colhead{Source} & \colhead{$a$} & \colhead{$b$} & \colhead{$c$} & \colhead{$n$}}
\startdata
\citet{Barnes 2007} &0.7725 $\pm$ 0.011& 0.601 $\pm$ 0.024 & 0.40    & 0.5189 $\pm$ 0.0070\\
\citet{Mamajek 2008} &0.407 $\pm$ 0.021 & 0.325 $\pm$ 0.024 & 0.495 $\pm$ 0.010 & 0.566 $\pm$ 0.008\\
\citet{Meibom 2009} &0.770 $\pm$ 0.005 & 0.472 $\pm$ 0.027 & 0.553 $\pm$ 0.052 & 0.52\\
\enddata\label{table:Parameters Gyrochronology}
\end{deluxetable*}

Here, we consider gyrochonology relationships that follow the form of \citet{Barnes 2007}. Color (mass) and age are treated as independent effects and thus mathematically separable into two functions as shown below:
\begin{align}\label{eq:gyrochronology}
P(B-V,t)&=f(B-V)g(t) \\
f(B-V)&=a[(B-V)_0-c]^b \\
g(t)&=t^n
\end{align}
where the age $t$ is in Myr. Spin down is accomplished with a power law $g(t)$ and an index of $n=0.5$ corresponds to the \citet{Skumanich 1972} law. The polynomial $f(t)$ models the color dependence of the slow rotator sequence at fixed age. Since $f(B-V)$ lacks time dependence, any relationship of this form produces self-similar mass-rotation isochrones. Although many variations exist, for the purposes of this paper, we compare three published gyrochronology relationships, with parameters listed in Table \ref{table:Parameters Gyrochronology}. These gyrochronology relationships are typically calibrated to cluster data in the color range $0.5 \lesssim \mathrm{(B-V)}_0 \lesssim 1.4$; \citet{Mamajek 2008}'s relation is defined over the smaller range $0.5 < \mathrm{(B-V)}_0 < 0.9$. Both \citet{Barnes 2007} and \citet{Mamajek 2008} base the color-rotation relationship on many well studied clusters (e.g.\ $\alpha$ Per, the Pleiades, Hyades, etc.) and determine the index $n$ as the factor that reproduces the Sun's rotation rate at the solar color. \citet{Meibom 2009} fit the color dependence in a single cluster (M35), but retained the power law index $n$ from \citet{Barnes 2007}.

\begin{figure}
\begin{center}
\plotone{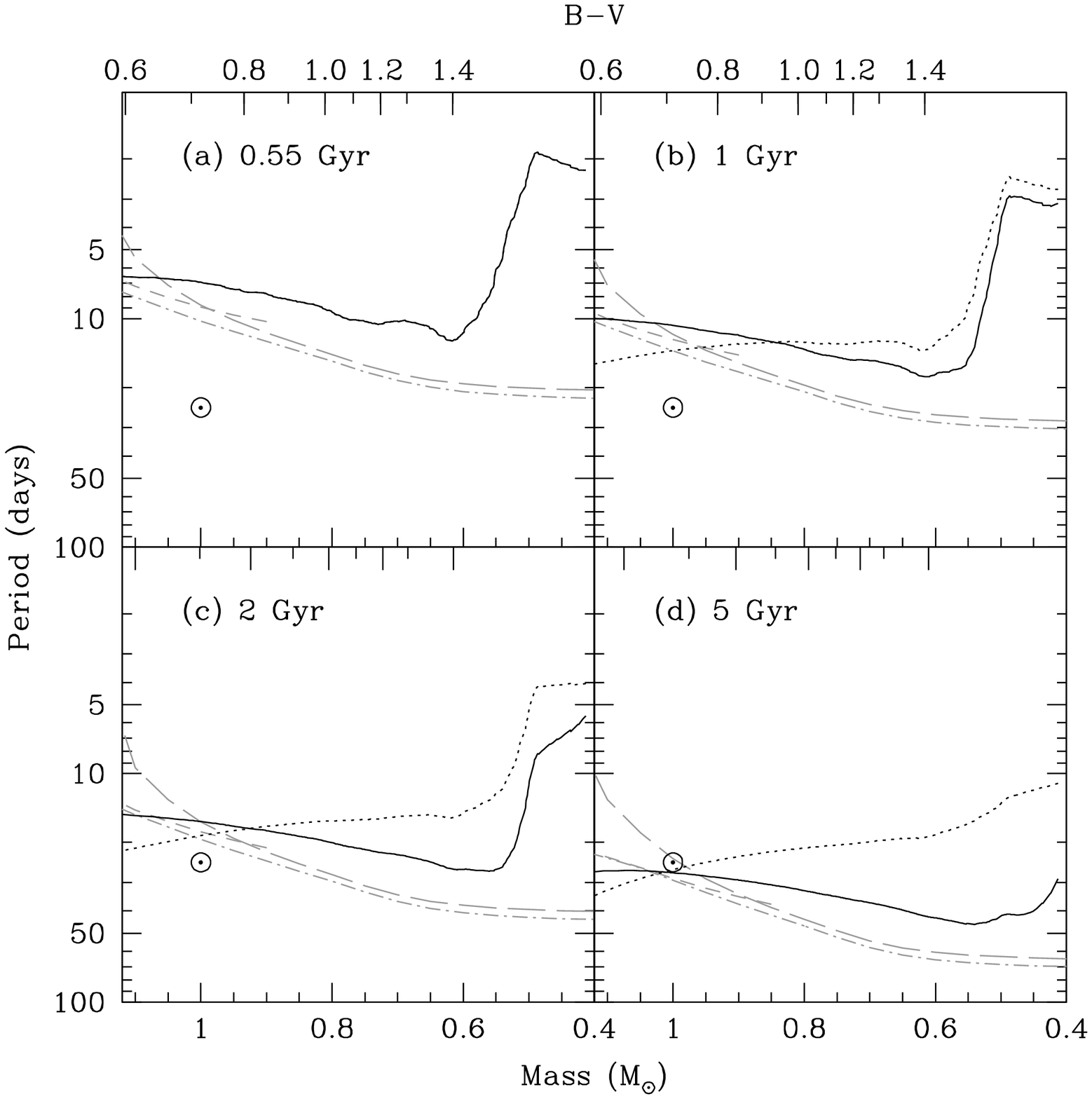}
\end{center}
\caption{Median mass-period relationship for a Kawaler (black solid) and \citeauthor{Reiners 2012} (black dotted) wind law at fixed age ranging from 0.55 Gyr (a) to 5 Gyr (d). Isochrones for three gyrochronology relations (\citealt{Barnes 2007}, dash-dot; \citealt{Mamajek 2008}, short dash; and \citealt{Meibom 2009}, long dash) are shown in gray for comparison. For reference, the Sun is shown at its current equatorial rotation period in all panels. The color-mass conversion used YREC isochrones of the respective age and [Fe/H]=+0.045.}\label{fig:gyrorange}
\end{figure}

Figure \ref{fig:gyrorange} shows the median mass-rotation relationship obtained by evolving the M37 rotation distribution forward with both a modified Kawaler and \citeauthor{Reiners 2012} wind law (from \S\ref{sec:timeEvolution}). It also features the predictions of three gyrochronology relations: \citet{Barnes 2007}, \citet{Mamajek 2008}, and \citet{Meibom 2009} (see Table \ref{table:Parameters Gyrochronology}.)

The medians show a steep upturn at low mass that is not present in the gyrochronology curves. This underscores a fundamental difference: the median is a measure of the entire distribution, while these published gyrochronology relationships are calibrated only to the slow rotators. A comparison between the median and gyrochronology relationships is fair in the converged regime because there the entire rotation distribution has collapsed to the narrow, slowly-rotating band fit by gyrochronology relationships. Figure \ref{fig:gyrorange} may be interpreted by assuming everything to the left of the upturn in the median trend falls in the converged regime (see \S\ref{sec:timeEvolution} for a more formal definition of the point of convergence.)

Although open clusters provide information about rotation in the young stars, the Sun is the only point used to calibrate angular momentum evolution models at ages older than 1 Gyr. Given its significance, we first check how well these models reproduce the Sun. For a visualization, panel (d) comes the closest in age to the Sun at 5 Gyr. By definition, the modified Kawaler and \cite{Reiners 2012} medians reproduce the Sun at the solar age. Two of the three gyrochronology relations also match the Sun's equatorial rotation rate within $\pm1$ day. The third, \citet{Meibom 2009}, predicts a rotation period $\sim20$ days for the Sun because they did not use it as a calibration point. Instead, \citet{Meibom 2009} retained the time-dependence from \citet{Barnes 2007} while changing the color-rotation function $f(B-V)$ to fit the open cluster M35. Nonetheless, the gyrochronology relations best agree with our median rotation-mass-age relations near solar color.

For stars redder than the Sun, all three gyrochronology relations spin stars down too quickly, resulting in predicted rotation periods that are too long by a factor of $1.5-2$ relative to a modified Kawaler spin down model. These published relations display zero-point differences relative not only to theory (panels b-d), but also to the observed rotation distribution of M37 (panel a). We note that this cluster's rotation distribution should reflect the true angular momentum content of its stars because the coupling timescale is shorter than the age of M37 \citep{Denissenkov 2010}. Gyrochronology relationships may predict rotation rates that are too slow because they are calibrated to younger clusters with stars whose core and envelope have not yet re-coupled. If the surface rotation rate masks a reserve of angular momentum in the core, gyrochronology relations would propagate forward the superficial rotation rate and predict artificially slow rotation at late times.

\citet{Mamajek 2008} noticed that the \citet{Barnes 2007} relation predicted rotation periods 50\% longer than those found in the Hyades. To address this, \citet{Mamajek 2008} carefully selected only stars that clearly lie on the converged sequence to include in their fit to the mass-rotation function $f(B-V)$. As a result, the \citeauthor{Mamajek 2008} isochrones reproduce the shape of the mass-rotation sequence better than other gyrochronology prescriptions. Differences between the shape of the gyrochronology curves and the median of the M37 data in panel (a) arise strictly from problems in the calibration of gyrochronology relations in young and intermediate age clusters. However, additional time-dependent differences can also arise because our angular momentum models evolve in time and the gyrochronology relations remain self-similar. Two effects in particular can impact the rotation distribution and therefore its median: evolutionary changes to stellar structure and magnetic saturation. The former is most important for the rapidly evolving, relatively high mass stars. The prescriptions for $\dot{J}$ also depend on the mass and radius. This structural dependence is much more significant for the \citeauthor{Reiners 2012} law than for the modified Kawaler law. Magnetic saturation is important for both loss laws. In the case of the modified Kawaler law, only low mass stars rotate faster than the saturation threshold at the age of M37 because of the Rossby-scaling (see Figure \ref{fig:M37movie} and Eq. \ref{eq:omegacrit}). In contrast, the saturation threshold is constant in the \citeauthor{Reiners 2012} law and affects stars of all masses because of the way it cuts through M37's rotation distribution (Figure \ref{fig:movieReiners}). Together these two effects can have a mild (modified Kawaler) to significant (\citeauthor{Reiners 2012}) time-dependent impact on the median rotation-mass-age relationship.

Two more recent papers (Barnes 2010; Barnes \& Kim 2010) have expanded the approach taken by \citet{Barnes 2007}.  Earlier gyrochronology work defined a rapidly rotating (C sequence) and a slowly rotating (I sequence) population based on open cluster data.  \citet{Barnes 2010} combines the two into a combined evolution prescription, while \citet{Barnes Kim 2010} considers the global applicability of the commonly adopted \citet{Kawaler 1988} angular momentum loss law.

\citeauthor{Barnes Kim 2010} raise an interesting point about Kawaler-style angular momentum loss laws: the predicted mass dependence in the torque at fixed rotation rate is small, and therefore one would not naturally produce the observed mass-dependent mean trend from a uniform set of initial conditions.  However, on the cool end this can simply be a product of saturated angular momentum loss for rapid rotators, while on the hot end the Kawaler formulation clearly must break down for stars with thin surface convection zones.  Such stars are not observed to spin down on the main sequence, and there is also a clear difference in the distribution of surface rotation rates between high and low mass stars even in young systems such as the Pleiades.

\citet{Barnes 2010} considers a description of rotation as a function of time with two terms, claimed to be empirically determined, and also considers some effects of a range of initial conditions.  There are interesting insights, including an estimate of the dispersion in age for solar mass stars arising from his preferred prescription and initial conditions, and some discussion of the transition from the high to the low branches of the cluster sequences.  In particular, the asymptotic age range determined for solar-mass stars is particularly relevant to this work (see \S\ref{sec:distwidth}). However, we have two significant reservations about the underlying approach.

The predicted time evolution is not explicitly coupled to physical models of stellar evolution, internal angular momentum transport, and angular momentum loss.  As a result, there are physical phenomena (such as core-envelope decoupling) which would have a strong impact on behavior and which are not included in the phenomenology.  There is also no cross-check on whether the implied torque from the model parameters is compatible with empirical constraints.

In principle one could, of course, construct purely empirical rotation-age-color relationships.   However, the \citet{Barnes 2010} models are also not rigorously based on empirical grounds.  Ranges of main sequence rotation periods (from 0.12 days to 3.4 days) are adopted, and assumed to be mass-independent.  Clear mass-dependent trends in rotation appear in very young clusters, in particular in the well-sampled Pleiades and M35 data sets. Neither these systems nor the open cluster M37 are compared with predicted rotation periods as a function of color.  This is a serious omission, because in our view M37 is the best intermediate-aged open cluster data set and the cluster rotation pattern (Figure 9 in \citealt{Barnes 2010}) is quite different from that demonstrated for the similar-aged Hyades-Coma systems in his Figure 11. Had \citet{Barnes 2010} performed this comparison, he would have found that the data require models with an initial period that is a strong function of mass to reproduce the slope of M37's slow-rotator sequence. This model therefore requires a detailed confrontation of predictions with the cluster data.

\subsection{Uncertainties in Rotation-Based Ages}\label{sec:uncertainties}

Until stars lose the memory of their initial conditions, we cannot distinguish between a relatively slowly rotating young field star and a relatively rapidly rotating older field star at the same rotation period. This implies that there exists a minimum error in the mean rotation-mass-age relationship due to the intrinsic dispersion in stellar rotation rates. We quantify this minimum age uncertainty and evaluate in what mass and period regimes the rotation period becomes a good estimator of age for field stars (\S\ref{sec:ageatfixedP}).

Furthermore, rotation period measurements suffer from uncertainties. While instrumental and observational errors may be removed by longer baseline monitoring or higher precision photometry, the fact that stars do not have a unique surface rotation period presents an irreducible uncertainty (see \S\ref{sec:currentresults} for a possible exception). Seismic inversion (e.g. \citealt{Schou 1998}) and observations of sunspots at different latitudes on the Sun's surface (e.g.\ \citealt{Lustig 1983}) agree that the solar rotation rate is significantly faster at the equator than near the poles. Because we cannot know if we are measuring equatorial rotation rates for unresolved stars, this sets bounds on the precision of rotation period, and therefore age measurements using this technique (\S\ref{sec:minPeriod}). We also consider how different modeling choices influence our derived uncertainties in Section \ref{sec:other uncertainties}.

\subsubsection{Uncertainty Due to Initial Conditions}\label{sec:ageatfixedP}

For our base case, we adopt a modified \citet{Kawaler 1988} angular momentum loss model. We distill the M37 distribution into a mass-dependent median rotation period and use the interquartile region to measure of the period uncertainty due to the spread of initial rotation rates. We evolve the smoothed 25$^\mathrm{th}$ and 75$^\mathrm{th}$ percentile curves (defined in \S\ref{sec:timeEvolution}) forward in time until we reach either the age of the Galactic disk, which we take to be $t=10$ Gyr, or when a model corresponding to that mass leaves the main sequence. Table \ref{table:gyrotable} contains the 10$^\mathrm{th}$, 25$^\mathrm{th}$, 50$^\mathrm{th}$, 75$^\mathrm{th}$, and 90$^\mathrm{th}$ percentile curves as a function of time, where (B--V)$_0$ and T$_\mathrm{eff}$ are computed at fixed mass using a YREC isochrone of the appropriate age and [Fe/H]=+0.045. Interpolation on this grid allows the determination of rotation-based ages and the associated uncertainty.

\begin{deluxetable*}{cccccccccccccc}
\tablecaption{Median Gyrochronology with Uncertainties for a Modified Kawaler and \citeauthor{Reiners 2012} Wind Law}
\tablehead{
\colhead{} & \colhead{} & \colhead{} & \colhead{} & \multicolumn{5}{c}{Modified Kawaler Wind Law} & \multicolumn{5}{c}{\citeauthor{Reiners 2012} Wind Law}\\
\colhead{} & \colhead{} & \colhead{} & \colhead{} & \multicolumn{5}{c}{-----------------------------------------------------} & \multicolumn{5}{c}{-----------------------------------------------------} \\
\colhead{Age} & \colhead{Mass} & \colhead{(B--V)$_0$} & \colhead{T$_\mathrm{eff}$} & \colhead{P$_{10}$} & \colhead{P$_{25}$} & \colhead{P$_{50}$} & \colhead{P$_{75}$} & \colhead{P$_{90}$} & \colhead{P$_{10}$} & \colhead{P$_{25}$} & \colhead{P$_{50}$} & \colhead{P$_{75}$} & \colhead{P$_{90}$} \\
\colhead{(Gyr)} & \colhead{(M$_\odot$)} & \colhead{(mag)} & \colhead{(K)}& \colhead{(days)} & \colhead{(days)} & \colhead{(days)} & \colhead{(days)} & \colhead{(days)} & \colhead{(days)} & \colhead{(days)} & \colhead{(days)} & \colhead{(days)} & \colhead{(days)}}
\startdata
0.55	&	0.45	&	1.54	&	3643	&	0.58	&	0.77	&	2.09	&	9.43	&	14.42	&	0.58	&	0.77	&	2.09	&	9.43	&	14.42	\\
0.58	&	0.45	&	1.54	&	3644	&	0.59	&	0.79	&	2.14	&	9.65	&	14.77	&	0.58	&	0.78	&	2.12	&	9.51	&	14.47	\\
0.63	&	0.45	&	1.54	&	3643	&	0.62	&	0.83	&	2.25	&	10.15	&	15.53	&	0.60	&	0.80	&	2.18	&	9.69	&	14.56	\\
0.69	&	0.45	&	1.54	&	3642	&	0.66	&	0.87	&	2.38	&	10.71	&	16.39	&	0.62	&	0.82	&	2.25	&	9.87	&	14.65	\\
0.76	&	0.45	&	1.54	&	3642	&	0.69	&	0.92	&	2.52	&	11.35	&	17.36	&	0.64	&	0.85	&	2.32	&	10.05	&	14.74	\\

$\vdots$ & $\vdots$ & $\vdots$ & $\vdots$ & $\vdots$ & $\vdots$ & $\vdots$ & $\vdots$ & $\vdots$ & $\vdots$ & $\vdots$ & $\vdots$ & $\vdots$ & $\vdots$ \\
\enddata\label{table:gyrotable}
\tablecomments{Table 2 is published in its entirety in the electronic edition of this journal. A portion is shown here for guidance regarding its form and content.}
\end{deluxetable*}

\begin{figure}
\plotone{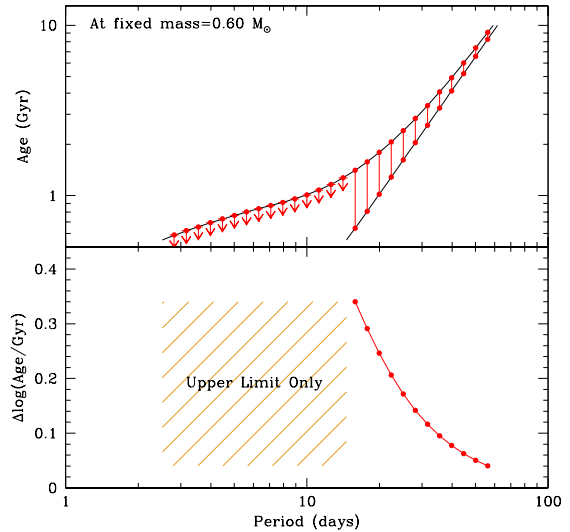}
\caption{\textit{Top:} Interquartile range of periods from M37 evolved forward using a solar calibrated Kawaler law for a bin of 50 stars centered on 0.60 M$_\odot$ (black lines). Vertical red lines cut the rotation profile at fixed period showing range of possible ages. Arrows indicate that gyrochronology can place either an upper limit on the age of a star with this period, but that it may be younger than 550 Myr; or, a lower limit, bounded by the age of the Galaxy. \textit{Bottom:} Width of interquartile band as a function of period. The quantity $\Delta\log(\mathrm{Age})$ gives the uncertainty in rotation-based age.}\label{fig:agewidth}
\end{figure}

\begin{figure}
\begin{center}
\plotone{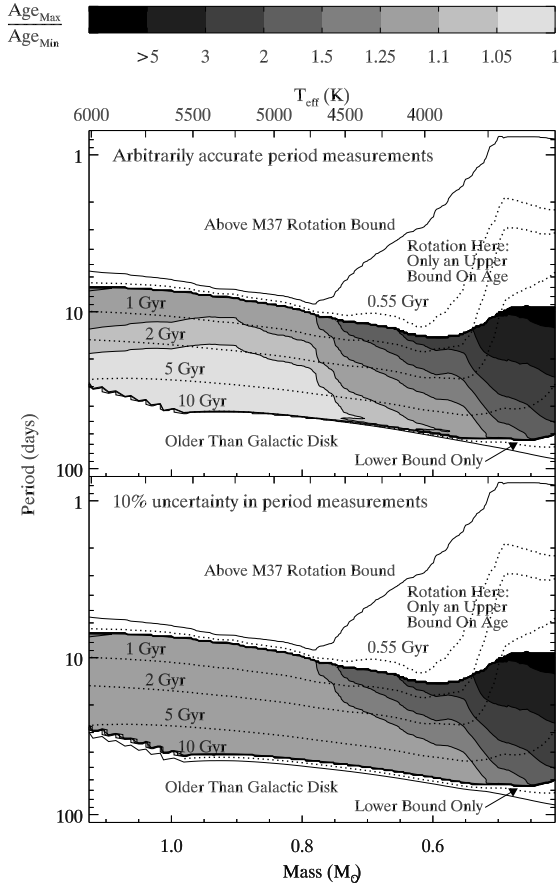}
\end{center}
\caption{Uncertainty in age ($\Delta\log(\mathrm{Age/Gyr})$) as a function of measured period and mass. The shaded region shows where gyrochronology is valid for our models. Uncertainties in age are calculated in the same manner as for Fig. \ref{fig:agewidth}. Lighter grays correspond to better constraints on the rotation age. Dotted lines give the median rotation period as a function of mass when the M37 distribution is evolved to 0.55, 1, 2, 5, and 10 Gyr using a solar calibrated Kawaler wind law. The mass axis is converted to temperature above assuming a 1 Gyr, [Fe/H]$_\mathrm{M37}=+0.045$ YREC isochrone. Both the panels assume that the solar calibrated Kawaler wind law perfectly captures the physics of angular momentum loss in cool stars and that masses can be measured to arbitrary precision. The upper panel assumes that rotation periods can also be perfectly measured, while the lower panel imposes a minimum 10\% uncertainty in the period measurement.} \label{fig:massrotationcontour}
\end{figure}

\begin{figure}
\begin{center}
\plotone{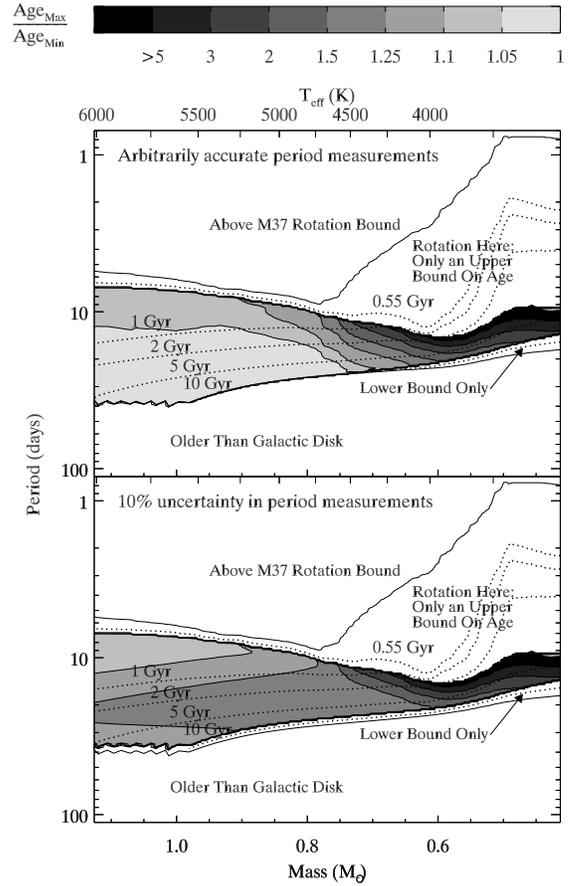}
\end{center}
\caption{Uncertainty in age ($\Delta\log(\mathrm{Age/Gyr})$) as a function of measured period and mass, where the initial mass distribution was calculated using YREC isochrones and evolved forward in time with the \citet{Reiners 2012} braking law. The upper panel assumes that rotation periods can also be perfectly measured, while the lower panel imposes a minimum 10\% uncertainty in the period measurement. See Figure \ref{fig:massrotationcontour} for an explanation of other markings.}\label{fig:Reinerscontours}
\end{figure}

The top panel of Figure \ref{fig:agewidth} shows an example of how the interquartile range changes over time at fixed mass. A horizontal cut across these curves gives the 25$^\mathrm{th}$ and 75$^\mathrm{th}$ period percentiles for a bin centered at M = $0.60$ M$_\odot$ at fixed age. In this calculation, the predicted range of rotation periods converges to a unique value at fixed mass and age. For field stars, it is useful to invert the problem and ask what range of ages is expected for a measured rotation period. This corresponds to taking a vertical cut through Figure \ref{fig:agewidth}, which gives the range in age predicted for a fixed rotation period. A unique rotation-mass-age relationship will predict a spurious range of ages for coeval stars born with different initial rotation rates because stars that are initially rotating faster take longer to reach a given period than initially slow rotators do. We take the difference between ages inferred from the top and bottom of our rotation distribution ($\Delta\log\mathrm{Age}$) as the uncertainty in the age estimate from gyrochronology (Figure \ref{fig:agewidth}). This ambiguity in age is a fundamental limitation of this technique; it takes a finite time for stars to forget their initial conditions. The lower panel shows the size of the age uncertainty for different measured rotation periods at fixed mass. The uncertainty in age declines toward longer periods because, if stars have time to spin down to long rotation periods, the rotation distribution will have had more time to converge.

There exists a region where gyrochronology can only provide an upper limit on the age of a star. We define this to be the case if the measured period lies within the interquartile range of M37 at 550 Myr. We lack further information due to our choice of initial conditions, and we discuss rotation-age distributions in young stars in a companion paper (Paper II). Only $\sim5\%$ of field stars should fall in the regime where only an upper limit is relevant, assuming the Galactic disk to be 10 Gyr old with a constant star formation rate. On the other end, at 10 Gyr, the period distribution still has some width and the slowly rotating quartile period sets a lower bound on the age and the upper bound is the age of the Galaxy.

Expanding our focus, we determine how the uncertainty in age depends not only on rotation period, but also on mass. In this ideal case, where we have assumed both perfect mass and rotation measurements, and that the theory fully encapsulates the physics of stellar angular momentum loss, the contours on the top panel of Figure \ref{fig:massrotationcontour} give the best possible constraints on age for each combination of mass and period, where lighter grays correspond to smaller $\Delta\log\mathrm{Age}$. The uncertainty in age increases towards lower mass because convergence takes longer for these stars. Gyrochronology can define an age range, rather than just place upper limits on the age, in the shaded region. In some windows of mass and period space, constraints tighter than $\Delta\log(\mathrm{Age})\sim0.05$ dex are possible, assuming perfect period measurements. As before, the region where gyrochronology can only provide an upper limit on the age arises from the initial spread in rotation rates. The lower bound occurs because we halted the simulation after 10 Gyr. The upward slope of the lower boundary at high mass represents these stars evolving off the main sequence. We offer the following cautions when interpreting this figure. First, because we use the interquartile region as our measure of the width of the rotation distribution, there will be stars that lie outside this range. Second, the specific values of the period and age uncertainty depend on the validity of the underlying angular momentum evolution model.

Thus far, we have used the widely-adopted \citet{Kawaler 1988} prescription for angular momentum loss, modified to include a Rossby-dependent saturation threshold as described in \S\ref{sec:kawaler}. While this simple model reproduces observations well in the open cluster domain; there is limited data for older field stars. Alternative formulations should therefore be explored, and the \citeauthor{Reiners 2012} prescription is an intriguing choice.

Following the same procedure as before, but instead using the \citeauthor{Reiners 2012} wind law produces the fractional age uncertainty contours displayed in Figure \ref{fig:Reinerscontours}. The \citeauthor{Reiners 2012} braking law introduces some interesting features that merit closer examination. Because the Reiners law enforces rapid rotation among low mass stars until late times, rotation provides only a weak constraint on age. For stars with M$>0.6$ M$_\odot$, rotation-based ages cannot be pinned down to better than 50\%, regardless of the precision of rotation period measurements. Among higher mass stars, \citeauthor{Reiners 2012} predict rapid angular momentum evolution until $\sim700$ Myr. If a star is caught with a rotation period in this range, its age may be determined to a precision of about 7\% (or approximately 30-50 Myr). Imposing a 10\% period measurement error (Figure \ref{fig:Reinerscontours}, bottom panel) produces a sweet spot for stars with rotation periods around 10 days.

\subsubsection{Impact of Latitudinal Differential Rotation and Period Errors}\label{sec:minPeriod}

Unfortunately, achieving these theoretical limits may not be feasible with real stars. Old field stars have a significant range in measured surface rotation rates which reflects latitudinal surface differential rotation (see \S\ref{sec:Old Field Star DR}). This provides an additional fundamental source of uncertainty in rotation-based ages: where on the surface is the rotation period being measured?

We broaden the range the ratio of quartile periods when it is smaller than the minimum 10$\%$ set by differential roation, centering $\Delta \log P$ on the geometric mean of the quartile rotation periods. That is,
\begin{align}\label{eq:differential rotation}
\Delta \log P&=\mathrm{max}\left(\log\frac{P_{75}}{P_{25}},2\times0.6745 \left[  \frac{ \sigma_P}{P\ln10}\right] \right) \\
\Delta \log P&=\mathrm{max}\left(\log\frac{P_{75}}{P_{25}}, \frac{1}{10 \ln10} \right).
\end{align}

We then repeat the process described above to obtain the plot shown in the bottom panel of Figure \ref{fig:massrotationcontour}. Comparison between the upper and lower panels shows that, even given a perfect angular momentum loss theory, latitudinal differential rotation at this level sets a floor on the age precision from gyrochronology estimates instead of allowing the range of rotation rates to converge arbitrarily.

For cool and young stars, the range in initial rotation rates dominates the error budget for rotation-based ages, but massive and older stars are instead limited by the period uncertainty set by differential rotation. For example, if a $0.5$M$_\odot$ star has a rotation period of 11.5 days, gyrochronology can only constrain its age to a factor of 4.7 regardless of whether there is a minimum precision in period. Meanwhile, the age of a 0.8 M$_\odot$ star with a period of 40 days can be tightly constrained to 2\% with perfect period measurements, but only to 20\% when we include a 10\% range due to differential rotation. For the Sun's mass and equatorial rotation period, its age can be constrained to 2\% with perfect period measurements. With this correction for differential rotation, the Sun's age is predicted to lie between 4.07 and 4.85 Gyr. Generally, imposing a 10\% range in rotation periods translates to a 20\% uncertainty in age.

\begin{figure}
\begin{center}
\plotone{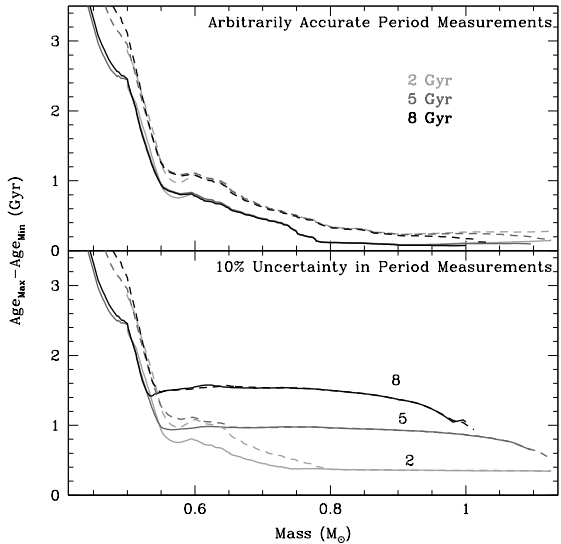}
\end{center}
\caption{Absolute age uncertainty as a function of mass for a star with the median rotation period at age 2, 5, and 8 Gyr (light to dark). Solid lines correspond to using the inner 50\% of stars to measure the distribution width while dashed lines indicate the inner 80\% of stars. The top panel assumes rotation periods are measured with arbitrary precision, while the bottom
panel includes a 10\% uncertainty in period measurements.} \label{fig:medianAgeError10-90}
\end{figure}

The dotted curves in Figure \ref{fig:massrotationcontour} delineate the median rotation period at a selected set of ages. At 550 Myr and 10 Gyr, rotation only sets a bound on the stellar age. At intermediate ages, the median rotation period may be mapped onto a typical uncertainty for a star of a given age. Although the logarithmic scale from Figure \ref{fig:massrotationcontour} is useful for visualizing the fractional error, the absolute age error also provides a helpful comparison. The solid lines in Figure \ref{fig:medianAgeError10-90} trace the absolute age uncertainties for stars with the median rotation period at select ages. Given arbitrarily accurate rotation period measurements, stars more massive than 0.75 M$_\odot$ always have an uncertainty in age smaller than 270 Myr. Adding in a 10\% uncertainty in rotation period has the effect of capping the age precision at around 0.38, 1, or 1.5 Gyr for a 2, 5, or 8 Gyr population, respectively. The absolute age uncertainties are smallest for younger stars, but become increasely larger for older stars. Nonetheless, the absolute age uncertainty is always below 2 Gyr for stars with M $> 0.52$ M$_\odot$. The wide binaries 61 Cyg and $\alpha$ Cen corroborate this expected level of uncertainty in rotation-based ages due to latitudinal surface differential rotation (\S\ref{sec:oldfieldstars}).

\begin{figure}
\begin{center}
\plotone{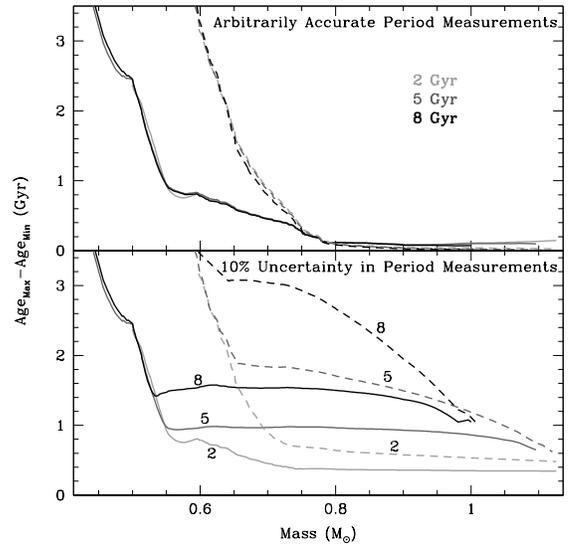}
\end{center}
\caption{Absolute age uncertainty as a function of mass for a star with the median rotation period at age 2, 5, and 8 Gyr (light to dark gray) calculated with Kawaler (solid) and \citeauthor{Reiners 2012} (dashed) braking law. The top panel assumes rotation periods are measured with arbitrary precision, while the bottom panel includes a 10\% uncertainty in period measurements.}\label{fig:Reinersmedian}
\end{figure}

Figure \ref{fig:Reinersmedian} shows the absolute age uncertainty predicted by both the Kawaler and \citeauthor{Reiners 2012} braking laws. When measurement uncertainty is introduced in the bottom panel, the \citeauthor{Reiners 2012} law predicts degraded precision in old stars because the \citeauthor{Reiners 2012} law has a weaker rate of angular momentum loss than the Kawaler law does at ages $>1$ Gyr (see Figure \ref{fig:reiners}). Both exhibit stable rotation-based age uncertainties with mass down to a critical point where the age uncertainty abruptly becomes quite large, corresponding to the point of convergence. This effective boundary to the rotation-based age technique is dramatically shifted 0.15 M$_\odot$ higher with the \citeauthor{Reiners 2012} law. This shift is due to the mass-independent way that \citeauthor{Reiners 2012} treat magnetic saturation. Comparison of Figures \ref{fig:M37movie} and \ref{fig:movieReiners} shows that \citeauthor{Reiners 2012} set the saturation threshold lower than the Rossby scaled saturation threshold for (B--V)$_0<1.4$ (M$\gtrsim0.57$ M$_\odot$). This keeps stars in the saturated regime longer than with a Rossby scaled saturation threshold. Because a Rossby-scaled saturation threshold is supported by observations of open clusters (see \ref{sec:kawaler}), we consider the Kawaler version to be a more realistic boundary to the rotation-based age technique.

Because the rotation-based age constraints differ strongly between these two angular momentum loss laws, properly testing the rate of angular momentum loss is important. For now, we favor the modified Kawaler wind law because it properly reproduces Skumanich spin down at late time for solar type stars and it incorporates a Rossby scaled wind law, both of which are supported by open cluster observations (see the discussion of Hyades and Praesepe in \S\ref{sec:Intermediate-AgeCluster}). Additionally, the \citeauthor{Reiners 2012} law fails to reproduce the young and intermediate age open cluster distribution, as seen in their Figure 4. They predict rotation periods that are too long compared to the cluster M50 for M$<0.8$ M$_\odot$. They argue that this is due to core-envelope decoupling, which is not included in their modeling. This can explain the presence of slow rotators over the mass range where core-envelope decoupling is expected to be important, primarily for solar-like stars with shallow surface convection zones.  However, low mass stars (below 0.5-0.6 M$_\odot$) are well described empirically with solid body spin down models \citep{Tinker 2002}. The striking differences between their models and cluster data for cool dwarfs therefore cannot be explained by this phenomenon. The modified Kawaler braking law developed in response to problems in reproducing empirical trends, and a comparable modification of the \citet{Reiners 2012} prescription could be viable and interesting to explore.

\subsubsection{Comparison with Prior Estimates of Age Uncertainties}\label{sec:distwidth}

We compare the absolute age uncertainties in Figure \ref{fig:medianAgeError10-90}. The largest differences occur for low mass stars, where the spread in rotation rates is maintained until late times. The metric chosen to describe the rotation width does not change the functional form. Even if arbitrarily accurate period measurements are possible, differences in the age uncertainty are only a few hundred Myr for stars on the tightly converged mass-rotation sequence. From the bottom panel, we see these differences due to the choice of width metric are smaller than the age uncertainty due to differential rotation for stars with M $>0.8$ M$_\odot$ at 2 Gyr and to even lower masses for older stars. We therefore consider the interquartile region a fair characterization of the rotation-based age uncertainty.

We compare our results with the age uncertainty calculated in \citet{Barnes 2010}. Instead of modeling the rotational evolution of the entire distribution, \citeauthor{Barnes 2010} abstracts the morphology of the rotation distribution into two sequences. \citet{Barnes 2010} defines an analytic equation whose limiting behavior matches the expressions proposed in \citet{Barnes Kim 2010} to reproduce the two sequences. Using this function, \citeauthor{Barnes 2010} finds that the absolute age range between the fastest and slowest rotators is a constant for a particular stellar mass, and is 255 Myr at a solar mass. To compare with our models, we adopt a more generous metric for the width of the distribution because \citealt{Barnes 2010} is measuring the slowest and fastest rotators in the distribution. At a solar mass, we predict the age range between the 10$^\mathrm{th}$ and 90$^\mathrm{th}$ percentile rotators is within $255\pm15$ Myr for the first 5 Gyr, in excellent agreement with the \citeauthor{Barnes 2010} prediction. This is different by a factor of $\sim2$ from the predictions that use the interquartile range to characterize the width for solar mass stars.

As can be seen from the top panel of Figure \ref{fig:medianAgeError10-90}, our models corroborate \citet{Barnes 2010}'s analytic result that the uncertainty in age is constant with time at a given mass when it is assumed that period measurements are arbitrarily accurate. Our results are more general for two reasons. We utilize open cluster data to capture the mass-dependent range of initial rotation rates rather than assuming qualitative estimates for starting conditions. We also adopt a physically-motivated model for angular momentum loss that incorporates evolutionary changes in stellar properties, like the moment of inertia.

\subsubsection{Impact of the Age and Mass Scale}\label{sec:other uncertainties}

Uncertainties in the absolute age assigned to M37 affect all of our derived rotation-based ages. In \S\ref{sec:initial conditions}, we initialize all models to the age of M37 determined from \citet{Hartman 2008}. By fitting isochrones to the cluster main sequence, \citet{Hartman 2008} measured an age of $550\pm30$ Myr. To see how this uncertainty in the initial conditions maps onto our age predictions, consider the illustrative case of Skumanich spin down. In that scenario, the rotation rate scales as $\omega\sim t^{-0.5}$. This implies that:

\begin{equation}
\frac{t_\bigstar}{t_\mathrm{M37}}=\left[ \frac{\omega_\mathrm{M37}}{\omega_{\bigstar}} \right]^2.
\end{equation}

Thus, the age of M37 ($t_\mathrm{M37}$) enters linearly into the calculation of a star's age ($t_\bigstar$). An analogous calculation may be performed for both the Kawaler and the \citeauthor{Reiners 2012} wind laws. As such, the $\pm$5\% uncertainty in the age of M37 \citep{Hartman 2008} propagates to a $\pm$5\% uncertainty in the ages predicted by these models. This uncertainty in the scale factor represents a systematic uncertainty for any semi-empirically calibrated model. The random error associated with the age of any particular cluster may be reduced by using many different clusters. However, the systematic uncertainty in the overall open cluster age scale will remain.

The wind law describes the angular momentum loss rate in terms of mass, rather than an observable quantity like color. Stellar interior models predict divergent mass-color relationships, particularly at the cooler end. Our age uncertainties depend only weakly for M $> 0.55$ M$_\odot$.

\subsection{Intermediate-Age Cluster Comparisons}\label{sec:Intermediate-AgeCluster}

The measured rotation distribution of open cluster stars serves as a snapshot of stellar angular momentum evolution at fixed age. If these clusters are indeed representatives of an underlying evolutionary sequence, then a successful model should be able to evolve one clusters' rotation distribution to match that of an older cluster. In this section, we show that this is the case. There are four intermediate-aged open clusters with substantial rotation samples and similar ages:  M37, Hyades, Praesepe, and NGC 6811. The first three clusters have reliable published ages: 550 Myr for M37 \citep{Hartman 2009} and 625 for Hyades and Praesepe (\citealt{Perryman 1998,Mermilliod 1981}.)

\begin{figure}
\begin{center}
\plotone{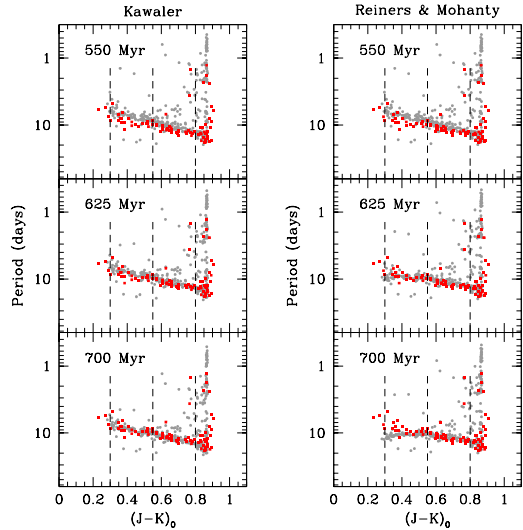}
\end{center}
\caption{Hyades rotation distribution (red) compared to M37's rotation distribution (gray). M37 is evolved forward from it's putative age of 550 Myr (top) forward in time to 625 Myr (middle; \citealt{Perryman 1998}) and 700 Myr (bottom; \citealt{Salaris 2004}). Two models of angular momentum evolution are considered: a modified Kawaler law (left) and \citeauthor{Reiners 2012} (right). The vertical dashed lines demarcate the color bins used to evaluate consistency.} \label{fig:Hyadesdist}
\end{figure}

\begin{figure}
\begin{center}
\plotone{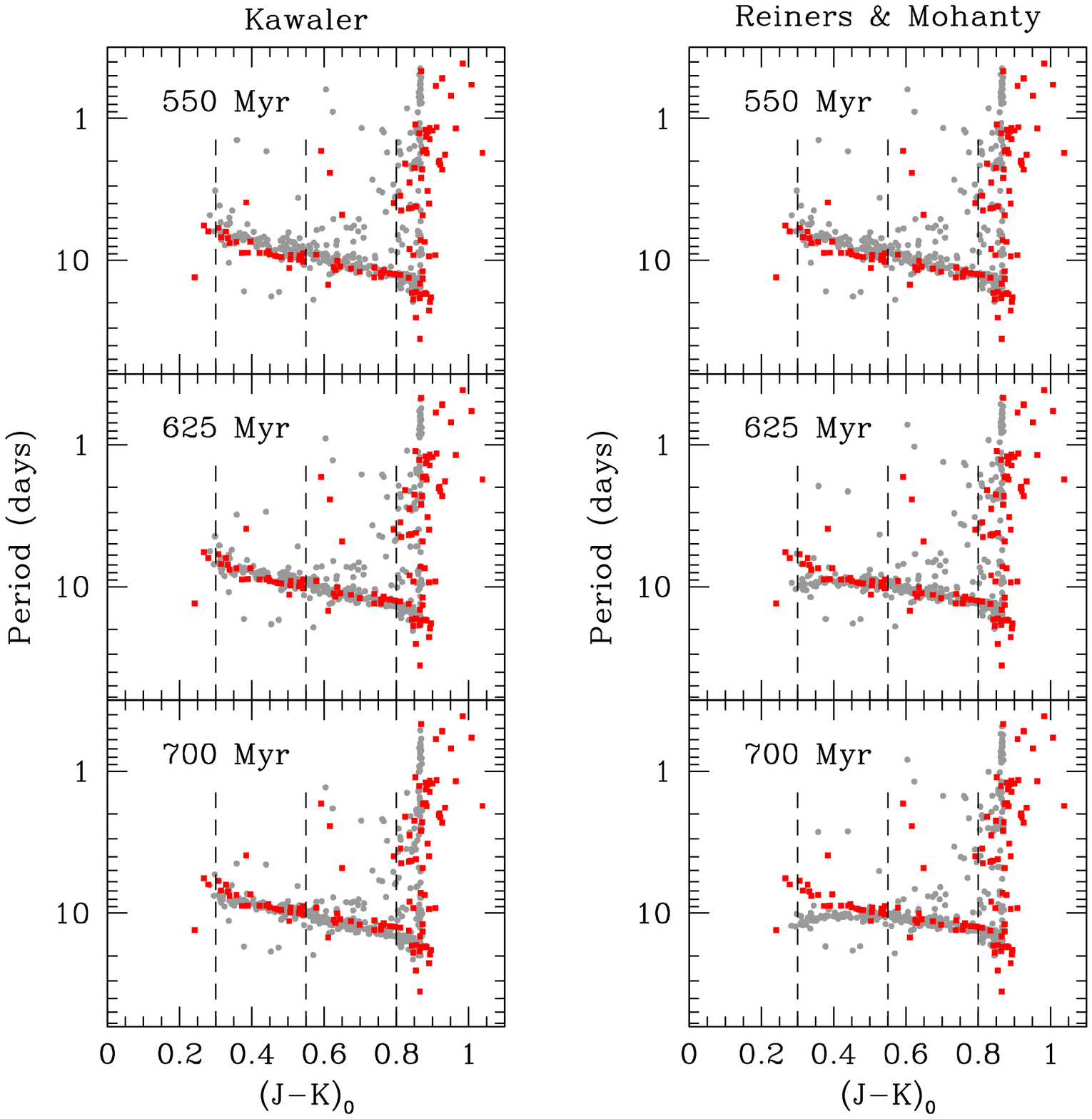}
\end{center}
\caption{Praesepe's rotation distribution (red) compared to M37's rotation distribution (gray). M37 is evolved forward from it's putative age of 550 Myr (top) forward in time to 625 Myr (middle; \citealt{Perryman 1998}) and 700 Myr (bottom; \citealt{Salaris 2004}). Two models of angular momentum evolution are considered: a modified Kawaler law (left) and \citeauthor{Reiners 2012} (right). The vertical dashed lines demarcate the color bins used to evaluate consistency.} \label{fig:Praesepedist}
\end{figure}

\begin{figure}
\begin{center}
\plotone{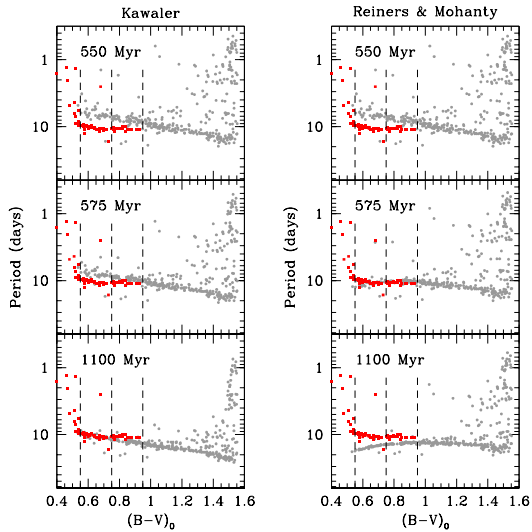}
\end{center}
\caption{A comparison of the NGC 6811 rotation distribution (red squares; \citealt{Meibom 2011b}) with the M37 rotation distribution (gray circles). The top panel shows the initial M37 distribution at 550 Myr \citep{Hartman 2009} while the lower panels show the M37 distribution evolved forward in time to the age displayed in the upper left. The left column corresponds to the modified Kawaler angular momentum loss law and the right column to the \citet{Reiners 2012} loss law. The vertical dashed lines demarcate the color bins used to evaluate consistency.} \label{fig:NGC6811}
\end{figure}

We begin with M37 at 550 Myr, the youngest cluster of the four clusters from \S\ref{sec:Intermediate-AgeClusterData}, and evolve it forward in time under two different models: a modified Kawaler prescription and the \citealt{Reiners 2012} wind law (\S\ref{sec:torque}). We use YREC isochrones to convert between mass and the appropriate color (B--V for NGC 6811 and J--K for Hyades and Praesepe) to compare M37's simulated future distribution with the other cluster data. Figures \ref{fig:Hyadesdist}, \ref{fig:Praesepedist}, and \ref{fig:NGC6811} depict the time evolution of M37 relative to the Hyades, Praesepe, and NGC 6811, respectively. The age depicted in the panels are chosen to reflect the fiducial age of M37 (550 Myr; top) and two of published age estimates for the comparison cluster described in \S\ref{sec:Intermediate-AgeClusterData}.

We focus on the regime where the rotation distribution has converged to a narrow band and divide this region into two color bins. For the Hyades and Praesepe, we choose a hot and cool bin with $0.3<(J-K)_0<0.55$ and $0.55<(J-K)_0<0.80$, respectively. For NGC 6811, we confine our analysis to the overlapping region of these two samples and choose bins $0.55<\mathrm{B-V}_0<0.75$ and $0.75<\mathrm{B-V}_0<0.95$.

\begin{figure}
\begin{center}
\plotone{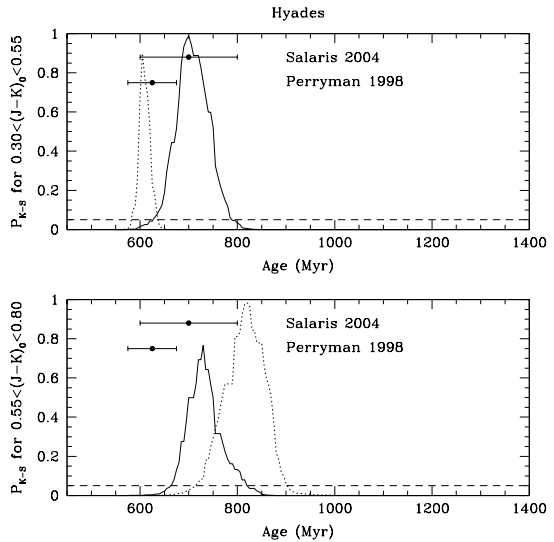}
\end{center}
\caption{The K-S probability that the Hyades rotation distribution is drawn from the same underlying distribution as the M37 distribution evolved forward in time to the age given in the abscissa. The probability is calculated using two models of angular momentum evolution: a modified Kawaler law (top) and \citeauthor{Reiners 2012} (bottom). The dashed horizontal line indicates the 5\% significance threshold. The points with errorbars depict selected age estimates for each cluster from the literature. We divide the converged rotation sequence in each cluster into a hot (solid) and cool (dotted) bin.} \label{fig:KSproball_Hyades}
\end{figure}

\begin{figure}
\begin{center}
\plotone{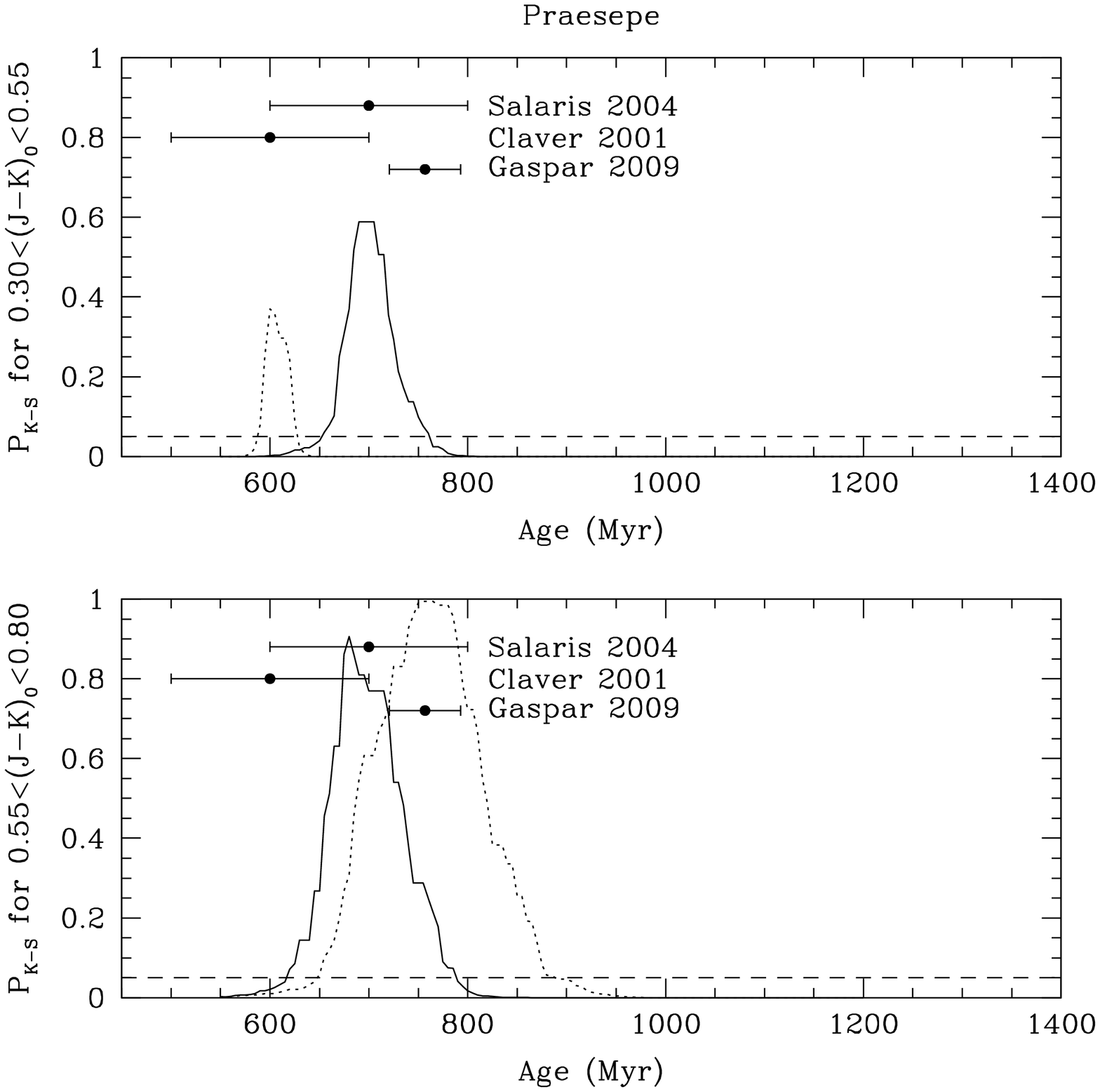}
\end{center}
\caption{The K-S probability that the Praesepe rotation distribution is drawn from the same underlying distribution as the M37 distribution evolved forward in time to the age given in the abscissa (see Figure \ref{fig:KSproball_Hyades} for details).} \label{fig:KSproball_Praesepe}
\end{figure}

\begin{figure}
\begin{center}
\plotone{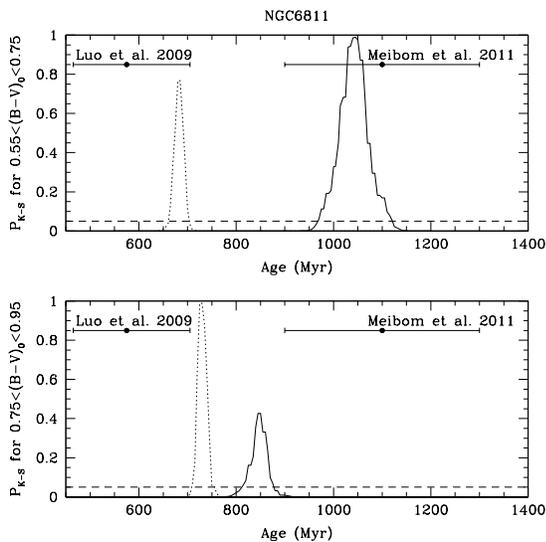}
\end{center}
\caption{The K-S probability that the NGC 6811 rotation distribution is drawn from the same underlying distribution as the M37 distribution evolved forward in time to the age given in the abscissa (see Figure \ref{fig:KSproball_Hyades} for details).} \label{fig:KSproball_NGC6811}
\end{figure}

Within each bin, we measure the agreement between the simulated M37 and observed cluster distributions by performing a Kolmogorov-Smirnov (K-S) test. Figures \ref{fig:KSproball_Hyades}, \ref{fig:KSproball_Praesepe}, and \ref{fig:KSproball_NGC6811} report the K-S probability as a function of the endpoint age of M37's simulated time evolution, broken down both by color bin and angular momentum loss prescription for the Hyades, Praesepe, and NGC 6811, respectively. We performed an additional test, where we allowed the initial age of M37 to vary by 10\%. This resulted in an overall shift to the absolute age scale but had no effect on the relative agreement between probability distributions in different color bins. We therefore limit the discussion to the case where M37's initial age is 550 Myr.

In the case of the Hyades, the region where the K-S probability exceeds 5\% overlaps between the hot and cool color bins for ages between 665 and 785 Myr with a modified Kawaler model. Not only are the hot and cool bins consistent with each other, but they also agree with both the \citet{Perryman 1998} and \citet{Salaris 2004} age estimates. In the \citeauthor{Reiners 2012} models, the hot and cool bins do not formally agree. The \citeauthor{Reiners 2012} model predicts rapid spindown among high mass stars, requires a younger age for bluer stars than for redder stars.

A similar pattern is seen with the Praesepe sample. The Kawaler models are self-consistent between the hot and cool bins and agree with isochrone age estimates. The \citeauthor{Reiners 2012} models differ relative to each other, but individually fall within the range of isochrone-based age estimates.

The opposite situation is seen for NGC6811. For this cluster, the \citeauthor{Reiners 2012} models are marginally self-consistent while the age predicted by the modified Kawaler law differs significantly in the hot and cool bins. Interestingly, the \citeauthor{Reiners 2012} models agree more strongly with the \citet{Luo 2009} young age for NGC6811 while the Kawaler model is more consistent with the older \citet{Meibom 2011b} age for the cluster.The rotation distribution of NGC6811 appears to flatline near a period of 12 days. Because the Kawaler models preserve the slope of the converged band in mass-period space, cool stars spindown down to a given period more quickly than hot stars do. The \citet{Reiners 2012} models actually invert the slope at the hot end, which agrees well with the flat observed NGC6811 distribution.

As noted by \citet{Meibom 2011b}, the observation that NGC 6811 rotation periods are flat with mass conflicts with data other old clusters (e.g.\ M37, Hyades, and Praesepe) that show a strong color dependence. This apparent inconsistency with other clusters could arise from crowding effects. Rotation period studies in the crowded cluster fields are challenging because Kepler's pixel scale is relatively large ($\sim4$''). For example, blending could cause a measured rotation period to be attributed to another star with a different color, smearing out a color-dependent sequence into a flat trend. This effect could explain why the modified Kawaler law, which is consistent with both the Hyades and Praesepe, fails to match this older cluster. Further rotation data, along with a robust age estimate, are needed for NGC 6811 to become a better calibration point.

Because M37, Hyades, and Praesepe are consistent under a modified Kawaler wind law, selecting any one of these clusters to initialize the models will yield the same behavior. Although it is possible to merge observations from these three clusters, such a heterogeneous sample would introduce complicated biases. It is therefore reasonable to select M37 to anchor our models of angular momentum evolution because it represents the largest and most homogeneous sample of the intermediate-aged clusters.

On the other hand, while the \citeauthor{Reiners 2012} wind law has problems reproducing the observed rotation distributions in young (see Paper II) and intermediate-aged (Hyades, Praesepe) clusters, it does produce marginal agreement with NGC 6811. This inconsistency with observations suggests that more consideration should be paid to the predicted mass-dependence of the \citeauthor{Reiners 2012} angular momentum evolution for stars younger than NGC 6811. However, the fact that the \citeauthor{Reiners 2012} wind law agrees with the oldest cluster for which we have measured rotation periods makes it interesting to compare its rotation predictions for older stars.

In summary, the Hyades, Praesepe, and M37 are consistent with one another for the Kawaler angular momentum loss law, but that the cool and hot subsets require different ages for the Reiners loss law. The situation is reversed for NGC 6811, but observational selection effects may be important there. We therefore conclude that using M37 as a template is reasonable.

\subsection{Tests in Old Field Stars and the Importance of Differential Rotation}\label{sec:oldfieldstars}

Older than 1 Gyr, the rotation-mass-age relationships described in \S\ref{sec:compare2gyrochronology} are constrained only by sparse data, primarily calibrated on the Sun. We are fortunate to possess rotation period and age measurements for two wide binaries. These old field stars can test whether the solar-calibrated rotation-mass-age relationship holds for stars with a mass and age different than the Sun.

Because stars lack a uniform surface rotation rate, any rotation period measurement of a star will be drawn from some distribution of surface rotation periods. If you measure a rotation period using a near-polar starspot, it will yield a different rotation period relative to an equatorial starspot because of latitudinal differential rotation (e.g.\ \citealt{Barnes 2005,Reiners 2006}.) The Sun exhibits differential rotation, with a rotation period of 25.4 days at the equator and 36 days around its poles (e.g. \citealt{Schou 1998,Beck 2000} and references therein). Disk integrated Ca II K-line measurements yield a range of rotation rates from 24.5 (lower than the true equatorial rotation rate due to measurement error) to 28.5 days, with a mean of 26.09 days \citep{Donahue 1996}.

The Sun is not alone in displaying differential rotation. Using Ca \textsc{II} H and K emission measurements from the Mount Wilson survey, \citet{Donahue 1996} report a range of rotation periods for 36 field stars with five or more seasons of detectable periodic modulation. \citeauthor{Donahue 1996} find a correlation between
$\Delta \mathrm{P}=\mathrm{P}_{max}-\mathrm{P}_{min}$ and $\left\langle \mathrm{P} \right\rangle$:
\begin{equation}\label{eq:Donahue}
\Delta\mathrm{P} \propto\left\langle \mathrm{P} \right\rangle ^{1.3\pm0.1}.
\end{equation}
If differential surface rotation is a generic stellar feature, we need to consider its effect when using old field stars to calibrate angular momentum loss models.

\subsubsection{Binary Star Data}

Binary stars with a measured rotation periods provide systems with measurable age that can constrain angular momentum loss at late times. Here we describe the data available for two wide binaries: 61 Cyg and $\alpha$ Cen.

\paragraph{61 Cyg}

Among these, the K5V+K7V binary 61 Cyg (HD201091 and HD201092) is of particular interest because it is physically well characterized. \citet{Donahue 1996} report mean rotation periods $\langle P\rangle_A=35.37$ days and $\langle P\rangle_B=37.84$ days with periods ranging from 26.82--45.25 days for 61 Cyg A and 31.78--46.57 days for B, measured over 12 and 13 observing seasons, respectively. We note that this observed range is larger than the typical for stars in the \citet{Donahue 1996} sample. The surface metallicity is spectroscopically measured as [Fe/H]$_A=-0.20\pm0.10$ and [Fe/H]$_B=-0.27\pm0.19$ \citep{Luck 2005}. The high-accuracy parallax information from \citet{Altena 1995} and the \emph{Hipparcos} catalog \citep{Perryman 1997} tightly constrains the luminosity to L$_A=0.153\pm0.010$ L$_\odot$ and L$_B=0.085\pm0.007$ L$_\odot$. Combining the parallax with interferometric radius measurements, \citet{Kervella 2008} found R$_A=0.665\pm0.005$ R$_\odot$ and R$_B=0.595\pm0.008$ R$_\odot$. Two measurements of the mass exist: 0.67 and 0.59 M$_\odot$ \citep{Walker 1995} and 0.74 and 0.46 M$_\odot$ \citep{Gorshanov 2006} for 61 Cyg A and B, respectively. \citet{Kervella 2008} used observational constraints to construct CESAM2k evolutionary models \citep{Morel 1997} and derive masses M$_A=$0.690 and M$_B=$0.605 M$_\odot$ and an age of $6\pm1$ Gyr.

We adopt the \citet{Kervella 2008}\ masses and re-derive the age with YREC models. We evolved models using the input physics described in \citet{van Saders 2012}, adopting a solar calibrated mixing length of $\alpha=1.926$ and Y = 0.265.  The initial metallicity was chosen to yield a surface [Fe/H] of $-0.20$ at an age of 6 Gyr.  We matched the radius within the errors at ages of $5.7\pm1.2$ and $7.5\pm2.5$ Gyr for 61 Cyg A and B respectively. This yields an age of $6\pm1.4$ Gyr for the system as a whole, to be compared with the age of $6\pm1$ Gyr derived using the CESAM code \citep{Kervella 2008} with the same masses and helium, but different choices of metallicity and mixing length.  However, we believe that this underestimates the age error, as an additional 1 Gyr systematic results from varying the metallicity with the observational error of 0.1 dex.  We therefore adopt $6.0\pm1.7$ Gyr as our age for the system.

\paragraph{$\alpha$ Cen}

The proximity of $\alpha$ Cen allows the precise determination of the fundamental parameters for its component G2V+K1V stars. We adopted the same physical parameters as \citet{Eggenberger 2004}. These include precise parallaxes \citep{Soderhjelm 1999}, masses \citep{Pourbaix 2002}, radii \citep{Kervella 2003}, and temperatures \citep{Neuforge 1997,Morel 2000}. Detections of p-mode oscillations in both $\alpha$ Cen A and B provide seismic constraints on the central helium abundance and age of the star \citep{Bouchy 2002,Carrier 2003}. Combining all of these non-seismic and seismic constraints, \citet{Miglio 2005} found an age of $6.8\pm0.5$ Gyr for $\alpha$ Cen; we adopt this age for $\alpha$ Cen. Asteroseismology is able to place much tighter constraints on the age of $\alpha$ Cen than is possible from comparison with isochrones, which range from 4 Gyr \citep{Demarque 1986} to $\sim7.6$ Gyr \citep{Guenther 2000}. The seismic age is consistent with isochrone-based age estimates. Although it is not used for age determination, we created a YREC model in agreement with the seismic parameters \citep{Eggenberger 2004,Miglio 2005} for stars of mass M$_A$=1.105, M$_B$=0.934 and Z = 0.0302, Y = 0.275 to properly model their interior structure for angular momentum evolution calculations.

Long-term monitoring of rotation rates from the Mount Wilson survey are not available because $\alpha$ Cen is not observable from the northern hemisphere. We refer to \citet{Mamajek 2008}, which cites two measurements of rotation period for $\alpha$ Cen A: $22\pm3$ days (Guinan 2008, private communication) and $28.8\pm2.5$ days \citep{Hallam 1991}. The rotation period of $\alpha$ Cen B is measured from a 1995 observing campaign in using the IUE satellite. Despite using different data reduction techniques, three groups obtain nearly identical results: $36.9\pm0.5$ days \citep{Jay 1997}, $35.1\pm$ days \citep{Buccino 2008}, and $36.2\pm1.4$ days \citep{DeWarf 2010}). As advocated by \citet{DeWarf 2010}, we adopt the average rotation period of $36.2\pm1.4$ days for $\alpha$ Cen B.

Because $\alpha$ Cen B has only one independent measurement of its rotation period, we adopt a range of rotation rates consistent with the average range observed in \citet{Donahue 1996}. We follow the formalism described in \ref{sec:minPeriod}. We center the adopted range about the single period measurement, where the range $\Delta\log \mathrm{P}$ is given by Eq. \ref{eq:differential rotation}.

\subsubsection{Surface Differential Rotation in Binaries}

\begin{figure}
\plotone{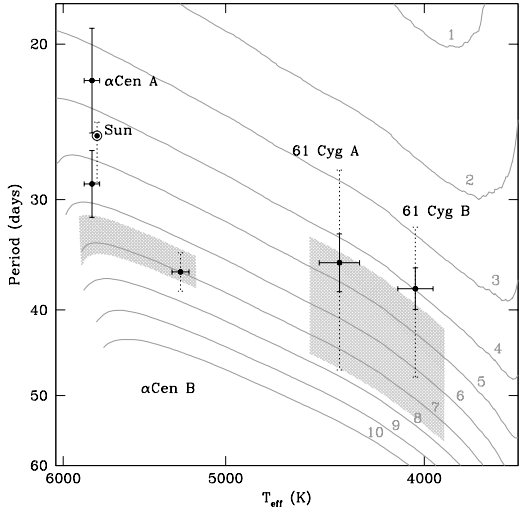}
\caption{Location of the Sun and wide binaries $\alpha$ Cen and 61 Cyg in the effective temperature-rotation period plane. The Mount Wilson survey measured the range of observed rotation periods over many seasons due to surface differential rotation (dashed), mean rotation rate (point), and error in the mean (solid) for the Sun and 61 Cyg \citep{Donahue 1996}. Because $\alpha$ Cen lacks such detailed rotation measurements, we show the two measurements available for $\alpha$ Cen A and assume that $\alpha$ Cen B displays a range in rotation rates similar to 61 Cyg A, which has a similar rotation period. The gray contours represent the median temperature-period relation for stars with age 1--10 Gyr, assuming a modified Kawaler wind law and YREC isochrones. The hatched regions illustrate the region consistent with the isochrone ages for the two binaries.} \label{fig:binaries-teff-per}
\end{figure}

Although individual rotation periods may be measured precisely, they are drawn from a range of rotation periods corresponding to the latitude of the active region on the stellar surface. The range of rotation periods observed for the Sun and 61 Cyg and inferred for $\alpha$ Cen are shown in Figure \ref{fig:binaries-teff-per}. These ranges are compared with the median rotation curves evolved forward in time from the M37 rotation distribution assuming a modified Kawaler wind law.

Because binaries form coevally, the rotation rates of individual components should lie along the same rotational isochrone. Furthermore, the shaded regions indicate the range of rotation periods consistent with the seismic age of $\alpha$ Cen A and isochrone age of 61 Cyg. The measured rotation periods for 61 Cyg A cover a broad range of ages from approximately 3--9 Gyr. The hatched region corresponds to the range of rotation periods expected for a star with age $6.0\pm1.7$ Gyr. Both components of 61 Cyg are consistent with the isochrone age and with each other.

For $\alpha$ Cen, the tightly constrained seismic age is consistent with the B component and the longer period measurement \citep{Hallam 1991} of A's rotation period. The short period measurement of $\alpha$ Cen A's rotation compared to the isochrone age demonstrates that any single measurement can have a $\sim2$ Gyr error in its rotation-based age due to surface differential rotation.

\begin{figure}
\plotone{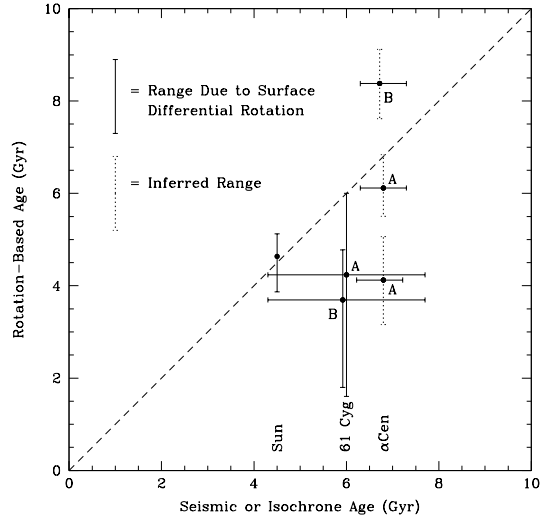}
\caption{Comparison of rotation-based age with isochrone age for the Sun and wide binaries $\alpha$ Cen and 61 Cyg. Rotation based ages calculated according to the median temperature-rotation relationship. As in Figure \ref{fig:binaries-teff-per}, the observed range of rotation rates are measured for the Sun and 61 Cyg. Since $\alpha$ Cen B has a single rotation period measurement, its range is assumed to be of comparable size to 61 Cyg A. Two individuals measurements of rotation period are shown for $\alpha$ Cen A, which together can be taken to set a minimum on its observed range of rotation rates. For clarity, the central age values of 61 Cyg B and $\alpha$ Cen B have been offset slightly from their true isochrone age.} \label{fig:binaries-ageage}
\end{figure}

To better quantify the range in rotation-based ages due to a range in observed rotation rates, we model the angular momentum evolution of these stars using a modified Kawaler wind law and the customized stellar evolutionary tracks described above. The initial rotation period is set to M37's median rotation period at the corresponding temperature and is evolved forward from 550 Myr. We use temperature as the coordinate to compare the near-solar metallicity M37 rotation distribution with the binaries, which have different metallicities and therefore different mass-color relationships, because the convective properties that govern angular momentum evolution are primarily set by T$_\mathrm{eff}$. We interpolate on this track to determine the rotation-based age corresponding to range of observed rotation rates. The results are compared with the seismic age of $\alpha$ Cen or the isochrone age of 61 Cyg as shown in Figure \ref{fig:binaries-ageage}.

Because these ages were computed with customized tracks, they will differ from the ages read off the near solar-metallicity M37 contours shown in Figure \ref{fig:binaries-teff-per}. In particular, this drives rotation-based ages for both $\alpha$ Cen B and the slightly metal-poor and lower mass 61 Cyg into worse (1$\sigma$--2$\sigma$) agreement with their respective isochrone ages, which hints at a possible metallicity dependence to the angular momentum loss law. Because the only point of calibration for loss laws at ages $\gtrsim1$ Gyr is the Sun, it is conceivable that stars of a different metallicity or mass would behave differently beyond the regime tested by clusters. The Kepler mission will yield a sample of field stars with asteroseismic ages capable of calibrating the metallicity dependence of the rotation-mass-age relationship.

We also perform an identical calculation using the \citet{Reiners 2012} wind law and the customized tracks. The \citeauthor{Reiners 2012} models do not spin down to periods longer than 30 days for $\alpha$ Cen B and 61 Cyg A and B within 10 Gyr. This model then fails to provide a reasonable rotation-based age for these binary components. \citet{Reiners 2012} note a related problem for field M stars, which are below the mass range considered here. However, even for K dwarfs slowly rotating stars exist with well measured ages younger than 10 Gyr. This indicates that the late time behavior of the \citeauthor{Reiners 2012} law as published does not appropriately capture true stellar behavior even for higher mass dwarfs.

If the \citeauthor{Reiners 2012} rotation-based ages are too old, some published gyrochronology relationships produce ages that are too young. We compare the two solar-calibrated published gyrochronology relationships against these binaries and determine that the \citet{Mamajek 2008} relation is consistent while the \citet{Barnes 2007} formalism is not. \citet{Mamajek 2008} already performed this comparison, finding an age (in years) of $\log t_A=9.57$ and $\log t_B=9.53$ of 61 Cyg's components using their mean rotation period, which average to $\langle \log t_\mathrm{61 Cyg} \rangle=9.55$. Although this mean age (3.52 Gyr) is too young compared with the isochrone age of 61 Cyg ($6.0\pm1.7$ Gyr), including the full range of rotation observed in 61 Cyg broadens the range of rotation-based ages to  $ 2.25<t_A< 5.66$ Gyr and $2.48<t_B<4.87$ Gyr. Similarly, for $\alpha$ Cen, \citet{Mamajek 2008} report $\langle \log t_{\alpha \mathrm{Cen}} \rangle=9.70$ (5.01 Gyr), which is inconsistent with the seismic age ($6.8\pm0.5$ Gyr) at the $3.6\sigma$ level if the gyroage is treated as exact. With the inferred range of rotation periods, the tension between the seismic age and the \citeauthor{Mamajek 2008} age ranges ($3.64 < t_A < 5.86$ Gyr and $4.68<t_B< 5.59$ Gyr) decreases to the $\sim2\sigma$ level. Thus, the mean \citeauthor{Mamajek 2008} trend is consistent with these binaries, but their estimated errors are too small. \citeauthor{Mamajek 2008} take the difference in predicted rotation-based ages of these coeval binary components as the error. They find an uncertainty of $\pm$11\% in gyrochronology predictions, similar to the total error of 15\% from \citet{Barnes 2007}. The analytic calculation of age uncertainty in \citeauthor{Barnes 2007} uses the \citet{Donahue 1996} relationship (see Equation \ref{eq:Donahue}) to estimate the period uncertainty due to surface differential rotation. We will discuss systematic sources of uncertainty in rotation-based ages, including surface differential rotation, in \S\ref{sec:uncertainties}.

We next consider the \citet{Barnes 2007} relationship. For conciseness, we write the age corresponding to the mean and range due to surface differential rotation in a condensed format. For 61 Cyg, the \citeauthor{Barnes 2007} relation yields ages of $t_A=2.52^{+1.53}_{-1.04}$ Gyr and $t_B=2.02^{+0.99}_{-0.58}$ Gyr. For $\alpha$ Cen, we find $t_A=4.67^{+1.19}_{-1.18}$ Gyr and $t_B=4.30^{+0.42}_{-0.40}$ Gyr. Even with the generous inferred range of surface rotation due to differential rotation, the \citeauthor{Barnes 2007} relation predicts ages that are systematically too young, especially in the case of $\alpha$ Cen B. This large discrepancy with old field binaries, coupled with its deviation from M37's median trend (Figure \ref{fig:gyrorange}), indicates that the \citet{Barnes 2007} relationship does not reproduce observations. Better rotation measurements for the $\alpha$ Cen system to determine a true, rather than inferred, range of differential rotation are important for testing rotation-mass-age relationships. These examples underscore the fact that differential rotation imposes an uncertainty in rotation-based ages and we will return to this point in \S\ref{sec:minPeriod}.

In addition to these binary systems, photometric monitoring programs have reported rotation periods for large samples of FGK (CoRoT, \citealt{Affer 2012}) and M (MEarth, \citealt{Irwin 2011}) field stars. Although these represent a heterogeneous stellar population that will include synchronized binaries as well as both dwarfs and subgiants, the rotation distribution gleaned by these studies still provides interesting constraints. These field star studies report rotation periods as long as 80--100 days for FGKM stars. The existence of such slowly spinning stars presents a challenge to current rotation-mass-age relationships, if they are main sequence stars. Because subgiants are expected to have longer rotation periods than dwarfs, we recommend a cautious interpretation of field star observations that do not distinguish evolutionary state. The three gyrochronology relationships tested here all predict that a 100 day period corresponds to ages older than 10 Gyr within the color regime B--V$<1.5$. \citet{Reiners 2012} recognize that the braking predicted by their wind law is much too weak to explain the slowly rotating M stars from \citet{Irwin 2011}. To account for this problem, they change their saturation threshold from 8.5 days to $P_\mathrm{crit}=40$ days below $M\leq0.3$ M$_\odot$. They justify this break in the saturation behavior with an argument that the effective $\tau_{CZ}$ rises sharply at low masses (e.g.\ \citealt{Stepien 1994,Kim 1996,Kiraga 2007}). With this adjustment, the \citeauthor{Reiners 2012} law reproduces the $\sim100$ day rotation periods for M stars, but retains rapid rotaion for more massive stars. Even with the adjusted $P_\mathrm{crit}$, the \citeauthor{Reiners 2012} law cannot explain the recent detection of equally long rotation periods in K and G dwarfs \citep{Affer 2012}. The modified Kawaler wind law also fails to predict 100 day periods. At 10 Gyr, our models predict a median rotation period of $\sim40$ days for G dwarfs, dropping to $\sim70$ days for M dwarfs. These long-period field stars will present a serious challenge to our understanding of angular momentum evolution if the period measurements prove to not be aliased and if further examination places these stars on the main sequence.

\section{Discussion and Conclusions}\label{sec:conclusions}

Our goal is to quantify the strengths and weaknesses of rotation as a stellar age indicator.  For which stars is rotation a good chronometer, and how reliable are those ages likely to be?  How does it compare to other age diagnostics? Both stellar rotation and activity diminish as stars age, and therefore rotation periods are harder to measure for old stars. How does the age limit for photometric surveys depend on the precision of the photometry?  In addition to these questions, there are broader issues that deserve comment.  Our data is concentrated in young stars, while the most promising applications are for old stars.  What potential issues could complicate the interpretation of rotation in older stars?  How can observational selection effects be accounted for?  And what backgrounds need to be understood for stellar population studies?  In our summary we begin with the quantifiable answers to the first class of issues, and then turn our attention to the challenges and opportunities for progress in the future.

\subsection{Current Results}\label{sec:currentresults}

Our main result is that we have placed quantitative limits on the precision of stellar rotation as an age-indicator for Sun-like stars. The upper mass limit for rotation-age relations is set by the F star transition from deep to shallow surface convection zones.  It is not clear that stars more massive than M $\simeq$ 1.2 M$_\odot$ have a rotation-age relationship, and it is difficult to detect rotation periods signals for stars more massive than the Sun because the fractional star spot coverage, and the amplitude of variability, drops steeply for hotter stars.

We demonstrated the method of deriving rotation-based ages from spin down models works for stars below the F star transition and above the fully convective limit by comparing the time evolved M37 rotation distribution with data from other intermediate-aged open clusters (\S\ref{sec:Intermediate-AgeCluster}). We demonstrate that M37, Hyades, and Praesepe are consistent under a modified Kawaler wind law, while the alternate \citet{Reiners 2012} braking law is able to reproduce the 1 Gyr old cluster NGC 6811. Although both theoretical angular momentum loss models and the semi-empirical published gyrochronology relationships are calibrated to the same open cluster data (M37), systematic errors arise in the extrapolation of these relationships forward in time. Because of the limited data at late times, constraining the rotational behavior of old stars is difficult. We find that both a modified Kawaler law and the \citet{Mamajek 2008} gyrochronology relation agree with constraints from old field binaries, although their predicted ages at fixed rotation period differ by as much as 30\% (\S\ref{sec:oldfieldstars}). We view this as a measure of the current systematic error level in rotation-age relations.

These loose constraints from old binary stars opens the door for other angular momentum loss laws offering a different spin down trajectory than the modified Kawaler law. A reassessment of the 25 year old Kawaler braking law in the modern context is welcome. However, the proposed \citet{Reiners 2012} is inconsistent with data. Taking the \citet{Reiners 2012} law at face value indicates that there are a strong mass- and radius-dependent trends in the spin down behavior at fixed $\omega_\mathrm{crit}$. The torque is predicted to be so inefficient that it conflicts with long period rotation measurements from the old binary star systems (61 Cyg and $\alpha$ Per), and field K and M dwarfs from CoRoT \citep{Affer 2012}, MEarth \citep{Irwin 2011}, and Mount Wilson survey (e.g.\ \citealt{Donahue 1996}). Future physically motivated revisions to the \citeauthor{Reiners 2012} law may solve these problems. The \citet{Barnes 2007} relation is also inconsistent with the binary star data. More rotation and age data is necessary for old stars to provide better guidance to theory.

The age errors for gyrochronology relationships are larger than optimistic early claims. We investigated the impact of the range of stellar birth rates (\S\ref{sec:ageatfixedP}) and surface latitudinal differential rotation (\S\ref{sec:minPeriod}) on the precision of age measurements.  These two phenomena will induce noise in age estimates of coeval stars even given perfect knowledge of the mean age-rotation relationship.  Below 0.6 -- 0.7 M$_\odot$, the minimum errors in rotational ages rise steeply; a range of surface rotation rates persists even for old stars.  The exact boundary depends on how the period data is transformed from the observational (color, period) domain into the theoretical (temperature, mass, age) domain.  Fortunately, there is good empirical guidance for both the time evolution of rotation and the proper choice of color-temperature relationships for low mass stars.  There is thus a window in the 0.6 -- 1.2 M$_\odot$ domain where stellar rotation is a promising chronometer.

Our technique for assigning errors for ages derived from rotation is conservative, and it may be possible to extract useful statistical age information from rotation in M dwarfs.  The first step would be to empirically measure the shape of the rotation distribution and to empirically calibrate the time dependence of the spin down in systems of known age, such as open clusters.  One could then construct a probability distribution of ages for a fixed rotation period.  In any given field, however, the true probability distribution would be a function of the age distribution in the background. A good example would be a field with a star cluster superposed on a sparse background. A blind method would assign a range of ages to an intrinsically peaked age distribution. The age distribution in a given field would therefore need to be obtained independently (for example, with rotation-based ages of K stars) and a star by star age estimate could then be constructed for a sample of M dwarfs for which age estimates are otherwise difficult. This method would also be particularly useful for determining the ages of special targets, like planet hosts.

Stars like the Sun do not have a unique surface rotation rate, and this injects another source of uncertainty in rotational age.  The latter effect becomes dominant in old field stars, imposing a minimum uncertainty of 20\% in age estimates for a 10\% range in surface rotation period.  Different asymptotic angular momentum loss prescriptions have a major impact on the limiting precision, with the recent \citeauthor{Reiners 2012} prescription implying much larger age uncertainties than previous spin down models based on a modified \citet{Kawaler 1988} loss law.  A proper empirical calibration of the late-time dependence should address the functional form of the rotation-age relationship, and it will also serve as a check on the width of the observed distribution. A proper calculation of the impact of surface differential rotation on period measurements would involve modeling of the number and latitude distribution of spots as a function of starspot cycle, and our method may overestimate the practical impact of surface differential rotation.  A larger problem may be the poor visibility of solar analogs during minima in their activity cycles (see below.)  Because the timescale for stellar activity cycles is of order decades, monitoring over many seasons (e.g.\ the Mount Wilson survey) would be necessary to measure a stars' full range in surface rotation periods. Nonetheless, surface differential is not a fundamental limitation in the same way as the initial conditions are. With extremely long term monitoring, surface differential rotation could in principle be calibrated out. In such a case, one might define a equatorial rotation period-mass-age relationship. However, the time investment to do this for a large number of calibrating stars is significant.

\subsection{Detectability of Rotation and Implications for Age Estimates with Upcoming Survyes}\label{sec:amplitudeage}

As demonstrated in the previous section, a regime exists where the range of stellar rotation rates has converged enough to make gyrochronology a good clock. For stars of the same mass, the age constraints from gyrochronology are tighter for more slowly rotating stars. Because magnetic activity, and therefore the presence of starspots, declines with time, so too does the amplitude of photometric variability. This raises the question of whether there exists a good mass window where stellar amplitudes of variability are high enough that current ground based surveys can obtain rotation periods for the older, less magnetically active stars that are the most promising candidates for gyrochronology. In the following discussion, we demonstrate that current ground based studies can measure rotation-based ages for stars up to 1-2 Gyr.

From the ground, current rotational studies have a minimum detectable amplitude of 3 millimag (e.g.\ \citealt{Hartman 2009,Meibom 2011a}), which may be pushed to as good as 1 millimag in the future. The space-based Kepler mission is monitoring roughly 150,000 stars with a time cadence of 30 min. Its photometric precision is estimated to be 1 millimag for a 15.5 mag star and as good as $10^{-5}$ mag for an 8$^\mathrm{th}$ mag star\footnote{http://keplergo.arc.nasa.gov/CalibrationSN.shtml}. Indeed, some studies of stellar amplitude from the first quarter of Kepler data (e.g. \citealt{Basri 2010,Basri 2011}) have already begun examining how the amplitude of variability depends on $T_\mathrm{eff}$, periodicity, and plan to investigate the age distribution of these stars once a longer time baseline is available.

In light of these upcoming developments, we use the current observations from open clusters to estimate what trends should be visible in the Kepler sample. To do so, we require a conversion between rotation period and amplitude. \citet{Hartman 2009} used the clean sample of rotation periods in M37 to explore the relationship between observed amplitude of variation and Rossby number. Dividing their sample at (B--V)$_0=1.36$, the blue stars display a strong anti-correlation between Rossby number ($R_0$) and amplitude which flattens out for $R_0<0.3$. Redward of that division, the anti-correlation is less pronounced and the constraints on the convective overturn timescale are weaker. Focusing on the bluer stars, \citeauthor{Hartman 2009} call into question the period measurements for stars with $R_0>0.6$, raising the possibility that they reflect spot evolution timescales rather than rotation periods. Based on this, \citeauthor{Hartman 2009} find an empirical fit to the r-band amplitude of M37 cluster members with (B--V)$_0<1.36$ and $R_0<0.6$ of the form
\begin{equation}\label{eq:HartmanR0Amplitude}
A_r=\frac{0.078\pm0.008}{1+\left(\frac{R_0}{0.31\pm0.02} \right)^{3.5\pm0.5}}.
\end{equation}

We invert Eq.\ \ref{eq:HartmanR0Amplitude} to determine the maximum observable rotation period for a survey of a given photometric precision. For some amplitude $A_r$, the resulting $R_0$ may be converted into a rotation period if the $\tau_{CZ}$ is known. We perform this calculation for a grid of different stellar masses where we utilize the $\tau_{CZ}$ predicted by the YREC models. Because $\tau_{CZ}$ changes only very slightly over a star's main sequence lifetime, we adopt a mass-$\tau_{CZ}$ relation at 1 Gyr. We restrict the grid to the range where the Hartman et al.\ relationship is applicable. To determine this, (B--V) is converted to mass using a 1 Gyr, [Fe/H]$_{M37}=+0.045$ YREC isochrone. Under this transformation, the color limit (B--V)$_0<1.36$ corresponds to masses M$>0.65$ M$_\odot$. This results in a mass-period relationship at fixed amplitude. These iso-amplitude curves represent the longest period for which a star of that mass could be detected with that photometric precision. A more slowly rotating star tends to have smaller variability in accordance with the activity-rotation-age relationship and therefore will drop below a given detection threshold.

\begin{figure}
\plotone{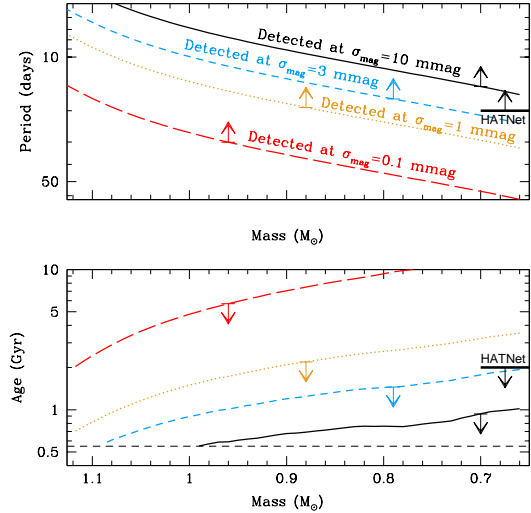}
\caption{\textit{Top:} Maximum observable rotation period for a survey capable of detecting amplitude variations of 10 (black; solid), 3 (black; short dash), 1 (gold; dot), and 0.1 (red; long-dash) millimag. \textit{Bottom:} The associate median age of a star with maximum period given in the top panel. Each line represents the maximum age a survey of a given precision could expect to obtain via gyrochronology. Arrows indicate the direction of the trends. The 550 Myr age bound is shown as the horizontal black dashed line. The HATNet survey (black bar) is capable of detecting the rotation periods of 20 days at amplitudes of 0.01 mag, which corresponds to ages of 2 Gyr or younger.}\label{fig:maxage}
\end{figure}

These lines of maximum period are shown in the top panel of Figure \ref{fig:maxage}. Because older stars rotate more slowly, if a survey is unable to detect periods longer than some threshold, this places an upper limit on the age of stars that can be determined with gyrochronology for that survey. We translate the maximum rotation periods into age using the median of the rotation distribution at a grid of ages given in Table \ref{table:gyrotable}. The mass range of these calculations is set by where Eq.\ \ref{eq:HartmanR0Amplitude} is applicable and by the scarcity of high mass stars in M37.

These maximum ages, shown in the bottom panel of Figure \ref{fig:maxage}, indicate that the current photometric precision of ground-based surveys (3 millimag) limits gyrochronology to measuring stars younger than $\sim1-2$ Gyr. An improvement of a factor of 3 in precision ($\sim1$ millimag) would open up the possibility of measuring rotation-based ages for stars as old as $2-3$ Gyr. A further order of magnitude improvement in precision is required to use gyrochronology to measure the ages of stars over the full age range of the Galactic disk (up to 10 Gyr). Kepler is capable of achieving the requisite precision to do so for stars brighter than m$_\mathrm{Kep}\sim$11 mag \citep{Basri 2010}.

The $R_0$-amplitude relationship suffers from a very limited dynamic range in $R_0$ from the M37 data. New observations of field stars will extend the relationship to larger $R_0$ (longer rotation periods) and provide a better constraint on the unsaturated slope. For now, we motivate a relationship with a shallower slope and consider its ramifications. This analysis is based on the \citeauthor{Hartman 2009} best fit curve that maps $R_0$ to amplitude. Some of the scatter in the M37 data is stochastic. The amplitude of variability of a star depends not only on the spot covering fraction, but also on the degree to which the spots are isotropically distributed over the surface of the star. At one extreme, if a star had a perfectly isotropic distribution of spots, no modulation in the light curve would be observed. Alternatively, large amplitude modulations would arise in the case of a concentration of spots along one line of longitude. By this logic, the upper envelope of the amplitude-$R_0$ diagram for M37 (see Figure 17, \citealt{Hartman 2009}) represents the maximum possible amplitude of variability for a star of a given $R_0$. The cloud of points under this envelope could be produced by different spot configurations that decrease the amplitude of variation. If instead we had used the upper envelope, rather than the best fit through the cloud of points, to define the the transformation from $R_0$ to amplitude, visual inspection indicates that the slope of the anti-correlation would be shallower. If this is the case, then using the best-fit $R_0$-amplitude relationship would represent an underestimate of the amplitude of variability. That is, there may be older, more slowly rotating stars which exhibit higher amplitude variability than one would expect from Eq.\ \ref{eq:HartmanR0Amplitude} because they possess a more favorable configuration of spots. This would enable ground-based surveys with sufficiently long time baselines to probe older ages than predicted in Figure \ref{fig:maxage}. Conversely, stellar variability may be suppressed to arbitrarily low levels by an nearly isotropic distribution of spots and therefore one should not expect all low amplitude stars to be old.

We can perform an empirical check on these maximum period-age-sensitivity predictions using current field star data. \citet{Hartman 2011} searched for stellar variability in light curves collected by the HATNet survey for transiting extrasolar planet. This yielded rotation periods for 1568 field F, G, K, and M dwarfs. HATNet offers a photometric precision of a few millimag for $R_C\sim 8$ mag and 0.01 mag for $R_C\sim12$ mag. Figure 18 in \citet{Hartman 2011} compares the mass-rotation distribution for field stars detected with HATNet with those from open clusters, including M37. At 0.7 M$_\odot$, the field stars have rotation periods a factor of $\sim2$ longer than the corresponding stars in M37. This also agrees with \citet{Hartman 2011}'s Figure 17 that shows that the number of detected stars rises slowly as a function of rotation period, peaks around 20 days, and then drops off steeply to larger periods for all color bins. If HATNet can detect rotation periods of $\sim20$ days or shorter, this corresponds to ages K-star ages 2 Gyr or younger from interpolation on Table \ref{table:gyrotable}. We overplot the HATNet detection capabilities in Figure \ref{fig:maxage} and discover that they more closely match the curve at $\sigma_\mathrm{mag}=0.003$ mag than the 0.01 mag curve that corresponds to the surveys detection threshold. This is an encouraging sign that our predicted detection limits are actually too pessimistic and the activity-rotation-age relationship does not fall off as steeply as Equation \ref{eq:HartmanR0Amplitude} predicts.

The results from HATNet mean that current groundbased surveys can identify populations of stars younger than 2 Gyr based on their amplitude of variability alone. However, such a sample could suffer contamination from merger products or synchronized binaries that are more active than isolated field stars of the same age. Additionally, the detection efficiency among slowly rotating is unknown. Stars inclined pole-on to our line of sight or those at the minimum in their stellar activity cycle contribute to a non-detection fraction of stars at a given period.

In summary, current large ground-based surveys can reach a precision of 0.01 mag, and targeted open cluster campaigns have demonstrated 0.003 mag precision levels.  We find that field K star populations younger than 1 and 2 Gyr (respectively) can be detected at these levels of precision if we adopt the amplitude-Rossby scaling relationship of \citet{Hartman 2009}.  The field star data of \citet{Hartman 2011} detected longer periods than would be predicted based on this estimate, suggesting that ground-based surveys may be capable of finding older stars than we predicted (2 Gyr at 0.7 M$_\odot$ and 0.01 mag precision, as opposed to our 1 Gyr estimate).  Much higher levels of precision, currently only possible with satellite data, are required to detect rotational modulation in old stars.  With new ground-based surveys on the horizon, such as LSST and the Las Cumbres Global Telescope Network, the monitoring large fractions of the sky will reveal variability for large samples of field stars. Because the amplitude of variability, like rotation rate, declines with age, such surveys will be able to identify all young K stars by variability alone.  A subset of active G stars will also be discovered, consisting of a very young ($<1$ Gyr) sample and tidally synchronized binaries.

\subsection{Cautions and Future Prospects}

Our method implicitly assumes that we can safely project the observational pattern seen in young systems into the future.  There are interesting theoretical classes of models where this is not the case.  For example, stars could vary in their core-envelope coupling because of differences in internal magnetic field morphology \citep{Charbonneau 1993}. This would lead to a persistent range in rotation rates at late ages generated by star-to-star differences in core rotation and the effective moment of inertia subject to spin down.  It may not be possible to maintain a dynamo if the rotation rate becomes too long relative to the convective overturn timescale \citep{Durney 1978}; both spin down and starspots could simply cease in stars above some critical age for a given mass.  Finally, gravitational settling could induce a composition gradient sufficient to decouple the surface convection zone from the interior.  There is some evidence from old open clusters \citep{Sestito 2005} that the pattern of lithium depletion stalls for systems older than a Gyr; if rotational mixing is tied to angular momentum loss \citep{Pinsonneault 1998a}, this could be evidence that there is a similar phenomenon in the torque.  These questions can be addressed with \textit{Kepler} data in old open clusters and targets with known seismic ages and measured rotation rates.

Interpreting rotation in field stars will also require careful consideration of observational selection effects and stellar population backgrounds.  It will be difficult to detect stars observed nearly pole-on \citep{Krishnamurthi 1998,Hartman 2010,Jackson 2010}. It will also be difficult to detect solar-like stars either at the minimum of their sunspot cycle or if they are in extended Maunder minimum-like epochs.  What is even more challenging will be modeling the stochastic effects of the distribution of star spots on the surface and observational selection effects from the observing cadence.  All of these effects conspire to produce a substantial population of non-detections, and careful work will need to be done to quantify their impact.

Finally, our age estimates, as well as the body of gyrochronology work, assumes that we have a sample consisting of main sequence single stars.  There will be significant background populations of tidally synchronized binaries and blue stragglers.  A few percent of solar-type stars are likely to be merger products \citep{Andronov 2006}, and their rotation rates will not be simple functions of their age.  Subgiants may be difficult to distinguish from dwarfs, especially for F and G stars, and their rotation-age relationships will be very different from one another.  All of these backgrounds will produce spurious signals when mapped onto single star relationships.  Finally, the metallicity dependence of spin down relationships is poorly understood because the calibrating open cluster sample has a very limited dynamic range in composition.

\subsection{Gyrochronology in the Context of Other Chronometers}

\begin{figure}
\plotone{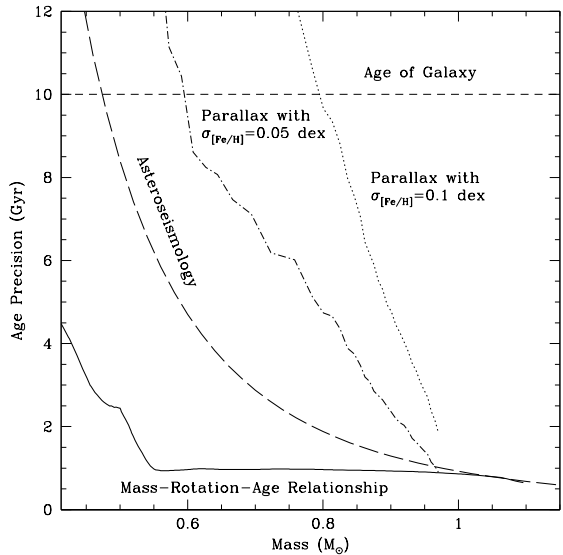}
\caption{Comparison of expected age precision for a 5 Gyr old star from different techniques of estimating the age of main sequence stars below the turnoff. Methods include rotation based ages (solid), asteroseismology (dashed), and trigonometric parallax combined with [Fe/H] measurements good to 0.1 dex (dotted) and 0.05 dex (dot-dash). }\label{fig:compareAgeTechniques}
\end{figure}

Rotation-based ages will be a welcome addition to the arsenal of age dating methods for unevolved dwarfs if they can be shown to be competitive with other techniques. The two other strongest contenders derive ages using constraints from parallax measurements or asteroseismology. For the former, ages may be determined by comparing a star's location on the CMD to stellar interior model predictions if the distance and metal content of a star is known. Gaia will provide exquisite geometric parallax measurements accurate to a few $\mu$as for $\sim10^9$ stars. However, without helium or metallicity information, stellar isochrones suffer a degeneracy with age. In contrast, asteroseismology obtains distance-independent age estimates for dwarfs by measuring the helium content. These ages are good to 10\% of a star's main sequence lifetime \citep{Kjeldsen 2009}, making the best age constraints possible for more massive stars.

We compare the expected age precision for these three techniques for a 5 Gyr old star as a function of mass in Figure \ref{fig:compareAgeTechniques}. We illustrate the sensitivity of parallax-based ages to metallicity by adopting 0.1 and 0.05 dex uncertainties in [Fe/H], optimistically assuming perfect distance, temperature, and helium measurements. To generate these curves, we use Dartmouth isochrones to determine what difference in age produces a change in luminosity equivalent to a shift of $\pm0.1$ dex (or 0.05 dex) in [Fe/H] at fixed temperature. Gaia's photometric accuracy will make the precision of metallicity information the limiting source of error. The asteroseismology curve is taken to be 10\% of the main sequence lifetime from the Dartmouth isochrones. The rotation-based age precision at 5 Gyr is taken from Figure \ref{fig:medianAgeError10-90}.

From this comparison, we see that asteroseismology potentially offers better age constraints than parallax-based techniques in the low mass regime. However, stellar pulsations may not even be detectable in low mass stars (e.g.\ \citealt{Chaplin 2011}). Because both geometric parallax and asteroseismology measure ages based on changes on the nuclear timescale, both suffer higher uncertainties in low mass stars. In comparison, Figure \ref{fig:compareAgeTechniques} demonstrates that rotation consistently offers ages good to $\sim1$ Gyr for stars M$>0.55$ M$_\odot$. Rotation-mass-age relationships are not competitive with other techniques for stars more massive than the Sun and the spread in rotation is large for low mass stars. Nonetheless, in an intermediate mass regime, rotation-based ages are not only competitive, but clearly the best possible age indicators available for field dwarfs.

\acknowledgements
We are grateful to Donald Terndrup for his thoughtful comments and illuminating discussions in the preparation of this manuscript. We would also like to thank an anonymous referees whose helpful comments improved the organization and clarity of this paper. C.E.\ and M.P.\ were supported by NSF grant AST-0807308 to The Ohio State University.

\end{document}